\documentclass[a4paper,fleqn,usenatbib]{mnras}
\usepackage{amsmath,amssymb}
\usepackage{graphicx}
\usepackage[table,dvipsnames]{xcolor}
\usepackage{booktabs}
\usepackage{multirow}
\usepackage{longtable}
\usepackage{float}
\usepackage{rotating}
\usepackage{gensymb}
\usepackage{xspace}
\usepackage{lineno}
\usepackage[nolist,nohyperlinks]{acronym}
\usepackage{hyperref}
\usepackage{natbib}
\hypersetup{colorlinks, citecolor=blue, linkcolor=blue, urlcolor=blue}
\newcommand{\angstrom}{\ensuremath{\mathring{\mathrm{A}}}}

\title[Compact Radio Emission in H\,0557--385]{H\,0557--385: A Candidate Missing Link Between Radio-Quiet Seyferts and Young Radio Galaxies}

\author[J. Chauhan et al.]{%
J.~Chauhan$^{1}$\thanks{E-mail: jc999@leicester.ac.uk}, T.~D.~Russell$^{2}$, A.~Lohfink$^{3}$, M.~Böttcher$^{4}$, F.~Panessa$^{5}$, R.~L.~C.~Starling$^{1}$,
\newauthor K.~Nyland$^{6}$, C.~Done$^{7}$, S.~Chattopadhyay$^{8}$, T.~Ghosh$^{8}$, R.~Misra$^{8}$, M. Del Santo$^{2}$\\
$^{1}$School of Physics and Astronomy, University of Leicester, University Rd, Leicester LE1 7RH, United Kingdom\\
$^{2}$INAF/IASF Palermo, via Ugo La Malfa 153, I-90146 Palermo, Italy\\
$^{3}$Department of Physics, Montana State University, P.O. Box 173840, Bozeman, MT 59717-3840, USA,\\
$^{4}$Centre for Space Research, North West University, Potchefstroom 2520, South Africa\\
$^{5}$INAF--Istituto di Astrofisica e Planetologia Spaziali, Via Fosso del Cavaliere, 00133, Roma, Italy\\
$^{6}$U.S. Naval Research Laboratory, 4555 Overlook Avenue Southwest, WA, DC 20375, USA\\
$^{7}$Centre for Extragalactic Astronomy, Department of Physics, Durham University, South Road, Durham DH1 3LE, UK\\
$^{8}$Inter-University Centre for Astronomy and Astrophysics, Ganeshkhind, Pune 411007, India,\\
}

\begin{document}
\label{firstpage}
\pagerange{\pageref{firstpage}--\pageref{lastpage}}
\maketitle

\begin{abstract}
The Seyfert\,1 galaxy H\,0557--385 ($z = 0.034$) has long been classified as radio-quiet on the basis of its radio-to-optical flux ratio, yet its radio properties are physically inconsistent with this designation. We present a broadband (0.076--258.8\,GHz) radio spectral analysis combining new Australia Telescope Compact Array (ATCA) observations with archival Murchison Widefield Array (MWA), Australian Square Kilometre Array Pathfinder (ASKAP) and Atacama Large Millimeter/submillimeter Array (ALMA) data. The spectrum requires a two-component model: a compact, self-absorbed component with a stratified gigahertz-peaked (GPS)/high-frequency-peaked (HFP) spectral shape (turnover at $9.4^{+0.6}_{-0.5}$\,GHz), superposed on an extended, optically thin power-law lobe ($\alpha_{\rm pl} = -0.78$) that dominates below ${\sim}1.6$\,GHz. The derived parameters --- a source size of ${\sim}15$\,pc, an equipartition magnetic field of 2.8\,mG, a synchrotron age of $\lesssim1.8\times10^{3}$\,yr, and a time-averaged jet power of ${\sim}5\times10^{41}$--$5\times10^{44}$\,erg\,s$^{-1}$ --- are characteristic of compact steep-spectrum and GPS young radio galaxies, not of canonical radio-quiet AGN. The two-component structure admits two interpretations that the present data cannot distinguish: the extended emission may be the growing lobe of a genuinely young, expanding radio galaxy, or relic plasma from an earlier episode of a transient or recurrent jet. Under either, H\,0557--385 is better understood as an intermediate object: optically a Seyfert, but with the radio-physical properties of a newly ignited or recently re-triggered radio galaxy. It is therefore a candidate missing link between radio-quiet AGN and large-scale radio galaxies. Distinguishing these scenarios will require high-resolution imaging and multi-epoch monitoring.
\end{abstract}


\section{Introduction}
The classification of active galactic nuclei (AGN) into radio-loud and radio-quiet categories has shaped our understanding of AGN physics for over half a century \citep[e.g.,][]{Kellermann1989, Sikora2007}. Radio-loud AGN are characterised by large-scale, relativistic jets that can extend to megaparsec scales, depositing enormous amounts of energy into the surrounding intergalactic medium and regulating the co-evolution of black holes and their host galaxies \citep{McNamara2007, Fabian2012, Morganti2017}. Radio-quiet AGN, by contrast, are thought to lack such prominent outflows; their radio emission is typically attributed to star formation, coronal activity, or weak, sub-relativistic winds \citep[e.g.,][]{Laor2008, Panessa2019}. This binary framework has provided a useful organisational structure, but its underlying classification has important limitations because it relies on radio-to-optical flux ratios measured at a single frequency---a metric that is sensitive to orientation, variability, host-galaxy contamination, and spectral shape \citep{Kellermann1989, Kellermann2016, Falcke1996, Padovani2017}. In its classical form, an AGN is deemed radio-loud if the ratio of its 5\,GHz to $B$-band (4400\,\AA) flux densities exceeds $R = S_{\rm 5\,GHz}/S_{4400\,\text{\AA}} > 10$, and radio-quiet otherwise \citep{Kellermann1989}.

Compact steep-spectrum (CSS), gigahertz-peaked spectrum (GPS), and high-frequency peaked (HFP) sources offer a useful counterpoint to the RL/RQ dichotomy. These young radio galaxies form a size sequence that is inversely related to their turnover frequency: HFP sources are the most compact (typically $\lesssim0.1$\,kpc, with turnovers above ${\sim}5$\,GHz), GPS sources are somewhat larger ($\lesssim1$\,kpc, turnovers near 1\,GHz), and CSS sources are the most extended (up to ${\sim}20$\,kpc, with turnovers at a few hundred MHz). Their radio spectra are characterised by a spectral turnover caused by synchrotron self-absorption (SSA) or free-free absorption (FFA) within the compact emitting region \citep[e.g.,][]{ODea1998, Orienti2014, Callingham2017}. The evolutionary sequence HFP $\rightarrow$ GPS $\rightarrow$ CSS $\rightarrow$ large-scale radio galaxy is well established for powerful, radio-loud systems \citep{ODea1998, Orienti2016}, yet the boundary between these young radio galaxies and the broader AGN population is not as clear-cut as once assumed. In particular, several objects with optical or X-ray properties that would classify them as Seyferts or low-luminosity AGN have been found to harbour GPS/CSS-like radio spectra, parsec-scale compact radio emission, and synchrotron ages of $10^4$--$10^6$\,yr \citep[e.g.,][]{Giroletti2009, Doi2013, Doi2016, Congiu2017, Inoue2018}.

A key limitation of the conventional RL/RQ classification is that it does not account for the possibility of young, newly ignited radio activity in sources that would otherwise be labelled radio-quiet. If a compact, self-absorbed synchrotron source is embedded in an AGN that is classified as radio-quiet on the basis of its radio-to-optical flux ratio \citep{Kellermann1989}, conventional radio-loudness criteria may substantially underestimate the true level of nuclear radio activity, because a low flux ratio reflects the relative optical brightness of the host and nucleus rather than the physical state of the radio-emitting plasma. Indeed, proper motions of mildly relativistic jets have now been measured via VLBI in systems such as Mrk\,110 \citep{Wang2025}, Mrk\,335 \citep{Yao2021}, and KISSR\,872 \citep{Kharb2024}, and the growing consensus is that the RL/RQ boundary is better understood as a continuum shaped by black hole spin, accretion rate, magnetic field topology, and evolutionary stage \citep{Blandford1977, Sikora2007, Merloni2003, Tchekhovskoy2010}, rather than a sharp physical divide. We note, however, that the detection of mildly relativistic ejecta is not by itself proof of a genuinely young radio source: episodic or recurrent nuclear activity can produce compact, relativistic outflows in an AGN that is not in the early stages of an HFP\,$\rightarrow$\,GPS\,$\rightarrow$\,CSS evolutionary track. Distinguishing genuine youth from transient activity requires additional evidence, such as a measured expansion rate or a steadily evolving turnover.

The radio spectrum provides the most physically direct window onto this question. A spectral turnover at GHz frequencies, whether produced by SSA or FFA, encodes the size, magnetic field strength, radiative age, and confinement state of the compact emitting region \citep{Kellermann1981, ODea1998, Orienti2014}.  For a source near the RL/RQ boundary, these parameters can distinguish a genuine young radio galaxy caught in the HFP or GPS phase of jet evolution, whose radio-to-optical flux ratio simply fails to reveal its radio-loud nature, from a transient, episodic synchrotron event in an otherwise radio-quiet system. Disentangling these scenarios requires broadband radio spectra spanning from hundreds of MHz to millimetre wavelengths, a capability that has only recently become available through facilities such as the MWA, ASKAP, ATCA, and ALMA operating in concert.

Here we present such a broadband spectral analysis of H\,0557--385 \citep[$z = 0.034$;][]{Fairall1982}, a Seyfert\,1 galaxy optically classified as radio-quiet whose radio properties are those of a young or recently re-triggered radio galaxy, placing it in an underexplored region between radio-quiet AGN and young radio galaxies. Adopting the nuclear optical continuum flux measured by \citet{Coffey2014} and our 5\,GHz flux density, the classical radio-loudness parameter is $R \approx 5$, placing H\,0557--385 formally in the radio-quiet regime ($R < 10$). We caution, however, that H\,0557--385 is a well-documented changing-look AGN whose nuclear emission varies by more than an order of magnitude between states \citep{Longinotti2009, Coffey2014}, so a single-epoch flux ratio is intrinsically unstable for this source; this sensitivity to the state of the nucleus is one of the limitations of the flux-ratio metric that motivates the radio-physical approach adopted here. Throughout this paper we use the phrase ``optically classified as radio-quiet'' to acknowledge that this designation rests entirely on radio-to-optical flux-ratio metrics.

The paper is structured as follows. Section~\ref{sec:Obs} presents the observations, data reduction, and imaging. Section~\ref{sec:jet_models} describes the broadband spectral analysis and modelling. Section~\ref{sec:physical_params} derives the physical properties. Section~\ref{sec:discussion} develops the missing-link interpretation, discusses the derived physical parameters and addresses the star-formation contribution. Section~\ref{sec:caveats} summarises the caveats and limitations, and Section~\ref{sec:conclusions} presents the conclusions and future prospects.

\section{Observations and Data Reduction}
\label{sec:Obs}
\subsection{ATCA Observations and Calibration}
\label{sec:ATCA}

We observed H\,0557--385 over two epochs in late November\,2022, under project code C3507.  Epoch\,1 was conducted from 2022 November\,25, 09:31:04.9\,UTC to 21:03:09.9\,UTC (MJD\,59908.3966--59908.8772), while Epoch\,2 ran from 2022 November\,28, 09:36:39.9\,UTC to 21:15:49.9\,UTC (MJD\,59911.4005--59911.8860). During both epochs, the ATCA was in a 6\,km extended configuration (6C\footnote{\url{https://www.narrabri.atnf.csiro.au/operations/array_configurations/configurations.html}}). For Epoch\,1, data were recorded simultaneously at central frequencies of 5.5\,GHz and 9.0\,GHz (C/X-band). For Epoch\,2, data were recorded simultaneously at central frequencies of 16.7\,GHz and 21.2\,GHz (K-band). Each frequency band had a bandwidth of 2\,GHz comprising 2048 1-MHz channels.

We used PKS\,B1934--638 for both bandpass and flux density calibration. The nearby calibrator PKS\,B0537--441, located 1.48$\degr$ from the target, was used for gain calibration. Data were calibrated using the Common Astronomy Software Applications package \citep[\textsc{casa} v6.6.0.20;][]{CASA_Team2022}, first being edited to remove data affected by radio frequency interference (RFI) or systematic artefacts, before being calibrated and imaged following standard \textsc{casa} procedures.\footnote{\url{https://casaguides.nrao.edu/index.php/Main_Page}}

\subsection{Imaging and Flux Density Measurements}
\label{sec:Imaging}

Imaging was performed using the \textsc{casa} task {\tt TCLEAN}. We adopted Briggs weighting with a robust parameter of 0.5 \citep{1995AAS...18711202B}, which offers a compromise between angular resolution and sensitivity while mitigating image artefacts from sidelobes. We used a circular restoring beam of $10\arcsec \times 10\arcsec$ with a position angle of 0\fdg0 east of north (a 9\,GHz ATCA continuum image of H\,0557--385 is presented in Fig.~\ref{fig:fig1}). Flux densities were measured by fitting for a source in the image plane using the {\tt IMFIT} task within \textsc{casa}; $1\sigma$ uncertainties were calculated by adding in quadrature the $1\sigma$ error on the Gaussian fit and the $1\sigma$ rms noise in the image.

We measured flux densities of $42.23 \pm 0.18$\,mJy at 5.5\,GHz, $41.90 \pm 0.24$\,mJy at 9.0\,GHz, $31.23 \pm 0.46$\,mJy at 16.7\,GHz, and $27.39 \pm 0.42$\,mJy at 21.2\,GHz. The ATCA data are not well represented by a single power law: the two-point spectral index between 5.5 and 9\,GHz is $\alpha = -0.016 \pm 0.015$, significantly flatter than the 9--21.2\,GHz index of $\alpha = -0.50 \pm 0.02$, suggesting a spectral break or turnover within, or close to, the ATCA 5.5--21.2\,GHz frequency coverage. To better constrain this feature, we divided the lower-frequency 2\,GHz sub-bands into three narrower bands, confirming a clear turnover in the radio spectrum (Fig.~\ref{fig:fig2}); see Section~\ref{sec:jet_models} for the detailed spectral analysis. The highest-frequency band, centred at 21.2\,GHz, was not sub-divided: it lies on the optically thin part of the spectrum, well above and away from the turnover, where finer frequency sampling adds no constraint on the turnover while the reduced bandwidth per sub-band would only inflate the per-point flux-density uncertainties.

To obtain comprehensive broadband coverage, we extend the radio spectrum to both lower and higher frequencies using archival observations from the MWA, ASKAP, and ALMA. The MWA flux densities were taken directly from the GLEAM-X catalogue \citep{HurleyWalker2022, Ross2024}, where source finding and deblending are handled by the survey pipeline. For ASKAP and ALMA we downloaded the images and measured the flux densities ourselves by fitting a single two-dimensional Gaussian to the source in the image plane, as described above. In every case the target is the only source detected within the field around H\,0557--385: visual inspection of each image showed no neighbouring source that could contaminate or blend with the target at the relevant resolution, so the measured flux densities are free of confusion from nearby sources. The full set of flux density measurements is listed in Table~\ref{tab:tab1}.

\begin{table*}
 \centering
 \caption{Details of the radio and millimetre-wave observations of H\,0557--385 used in this work.}
 \label{tab:tab1}
\begin{center}
\scalebox{1.0}{%
\begin{tabular}{|l|c|c|c|c|c|c|c|}
 \hline\hline
 Telescope & Observation  & Observation & Central   & Synthesized & Synthesized   & Flux  & References \\
           & date         & time        & frequency & beam size   & beam position & density$^{\mathrm{a}}$ &  \\
           & (DD-MM-YYYY) & (hh:mm:ss)  & (GHz)     & (arcsec)    & angle (deg.)  & $S_{0}$ (mJy) & \\
 \hline
 $\,$MWA & 2020-09-28 to 2020-10-25 & --- & 0.087 & $157.7\times133.5$ & $-78.3$ & $95.55\pm7.64$ & $[1]$ \\
 $\,$MWA & 2020-09-28 to 2020-10-25 & --- & 0.118 & $114.5\times98.1$ & $-79.2$ & $91.80\pm7.34$ & $[1]$ \\
 $\,$MWA & 2020-09-28 to 2020-10-25 & --- & 0.154 & $87.4\times77.1$ & $-80.1$ & $82.29\pm6.58$ & $[1]$ \\
 $\,$MWA & 2020-09-28 to 2020-10-25 & --- & 0.185 & $71.9\times63.9$ & $-82.1$ & $81.65\pm6.53$ & $[1]$ \\
 $\,$MWA & 2020-09-28 to 2020-10-25 & --- & 0.215 & $61.8\times54.5$ & $-81.7$ & $72.92\pm5.83$ & $[1]$ \\
 $\;$ASKAP & 2020-10-16 & 22:50:29 & 0.887 & $25.0\times25.0$ & 0.0 & $41.21\pm0.60$ & $[2]$ \\
 $\;$ASKAP & 2020-10-16 & 22:02:02 & 0.887 & $25.0\times25.0$ & 0.0 & $41.17\pm0.62$ & $[2]$ \\
 $\;$ASKAP & 2022-04-18 & 08:25:13 & 0.887 & $12.2\times11.6$ & 41.2 & $39.80\pm0.32$ & $[2]$ \\
 $\;$ASKAP & 2024-01-06 & 15:27:25 & 0.943 & $11.8\times10.2$ & 54.8 & $42.33\pm0.76$ & $[2]$ \\
 $\;$ASKAP & 2024-01-06 & 15:44:30 & 0.943 & $11.7\times10.2$ & 52.7 & $40.94\pm0.32$ & $[2]$ \\
 $\;$ASKAP$^{\dagger}$ & 2021-01-25 & 13:43:00 & 1.367 & $25.0\times25.0$ & 0.0 & $32.59\pm0.45$ & $[2]$ \\
 $\;$ASKAP$^{\dagger}$ & 2021-01-25 & 13:43:00 & 1.367 & $8.8\times8.0$ & 60.8 & $30.51\pm0.27$ & $[2]$ \\
 $\;$ASKAP & 2022-01-12 & 15:28:59 & 1.655 & $8.7\times6.6$ & $-53.7$ & $29.02\pm0.30$ & $[2]$ \\
 $\;$ATCA & 2022-11-25 & 09:47:04 &  4.68  & $10.0\times10.0$ & 0.0 & $40.06\pm0.14$ & This work \\
 $\;$ATCA & 2022-11-25 & 09:47:04 &  5.50  & $10.0\times10.0$ & 0.0 & $42.23\pm0.18$ & This work \\
 $\;$ATCA & 2022-11-25 & 09:47:04 &  6.32  & $10.0\times10.0$ & 0.0 & $43.45\pm0.18$ & This work \\
 $\;$ATCA & 2022-11-25 & 09:47:04 &  8.18  & $10.0\times10.0$ & 0.0 & $\phantom{1}43.32\pm0.30$ & This work \\
 $\;$ATCA & 2022-11-25 & 09:47:04 &  9.00  & $10.0\times10.0$ & 0.0 & $\phantom{1}41.90\pm0.24$ & This work \\
 $\;$ATCA & 2022-11-25 & 09:47:04 &  9.82  & $10.0\times10.0$ & 0.0 & $\phantom{1}41.37\pm0.30$ & This work \\
 $\;$ATCA & 2022-11-28 & 09:56:55 & 15.88  & $10.0\times10.0$ & 0.0 & $\phantom{1}33.09\pm0.58$ & This work \\
 $\;$ATCA & 2022-11-28 & 09:56:55 & 16.70  & $10.0\times10.0$ & 0.0 & $\phantom{1}31.23\pm0.46$ & This work \\
 $\;$ATCA & 2022-11-28 & 09:56:55 & 17.52  & $10.0\times10.0$ & 0.0 & $\phantom{1}30.50\pm1.08$ & This work \\
 $\;$ATCA & 2022-11-28 & 09:56:55 & 21.20  & $10.0\times10.0$ & 0.0 & $\phantom{1}27.39\pm0.42$ & This work \\
 $\;$ALMA & 2017-09-20 & 09:16:21 & 240.842 & $0.041\times0.032$ & 26.4 & $\phantom{2}2.80\pm0.04$ & $[3]$ \\
 $\;$ALMA & 2017-09-20 & 09:16:21 & 242.765 & $0.040\times0.030$ & 33.0 & $\phantom{2}2.71\pm0.03$ & $[3]$ \\
 $\;$ALMA & 2017-09-20 & 09:16:21 & 258.785 & $0.038\times0.030$ & 28.0 & $\phantom{2}2.56\pm0.04$ & $[3]$ \\
 \hline\hline
\end{tabular}}
\end{center}
\begin{flushleft}
$^{\mathrm{a}}$ $1\sigma$ errors calculated by adding in quadrature the $1\sigma$ error on the Gaussian fit and the $1\sigma$ rms noise in the image. These are measurement uncertainties only; the absolute flux-density calibration uncertainty is not included here but is incorporated in the spectral modelling (Section~\ref{sec:jet_models}), where a systematic uncertainty of 10 per cent for MWA and 5 per cent for ASKAP, ATCA, and ALMA \citep{Partridge2016} is added in quadrature to every point. \\
The synthesized beam size is the FWHM of the restoring (synthesized) beam, given as major\,$\times$\,minor axis. For the MWA (GLEAM-X) data the beam is that of the corresponding sub-band image and scales approximately as $1/\nu$, so it is largest at the lowest frequencies; the ATCA images use a common circular restoring beam. \\
$^{\dagger}$ These two observations correspond to two different ASKAP beams; the image in which the source lies closer to the beam centre has the smaller flux-density uncertainty.\\
$[1]$ -- \citet{HurleyWalker2022, Ross2024}, $[2]$ -- \citet{Duchesne2023, Duchesne2024, Duchesne2025} and $[3]$ -- ALMA project code 2016.1.01385.S\\
\end{flushleft}
\end{table*}

\subsection{Radio Morphology}
\label{sec:morphology}

\begin{figure}
    \includegraphics[width=\columnwidth]{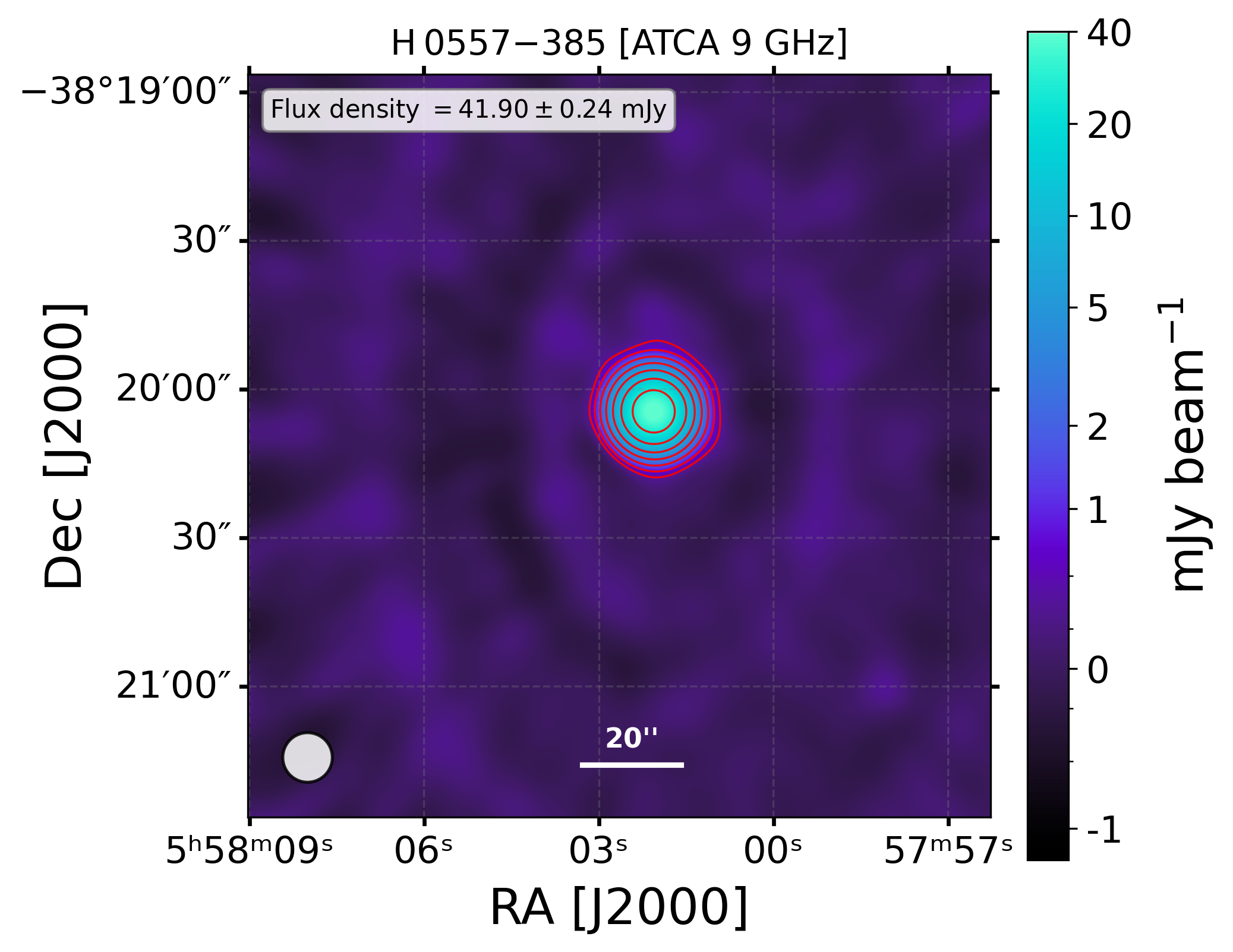}\vspace{-0.7cm}
    \caption{ATCA 9\,GHz continuum image of H\,0557--385 obtained in 2022 November. The image is centred on the target coordinates. The beam is circular with a FWHM of 10\,arcsec, shown in the lower-left corner. H\,0557--385 is unresolved, consistent with a compact morphology on kiloparsec scales. Contours are drawn starting from $3\sigma$ ($\sigma = 0.24$\,mJy\,beam$^{-1}$) and increasing by factors of two. The shallow negative ``bowl'' surrounding the source is a deconvolution artefact arising from the absence of short (zero-spacing) baselines in the interferometric $uv$-coverage. Because the flux densities are measured by fitting the source component directly in the image plane (Section~\ref{sec:Imaging}), they are unaffected by this large-scale negative feature.}
    \label{fig:fig1}
\end{figure}

Fig.~\ref{fig:fig1} shows the 9\,GHz ATCA continuum image of H\,0557--385 from 2022 November\,25. The source is detected with a peak significance exceeding $100\sigma$, with a peak flux density of ${\approx}42$\,mJy at both 5.5 and 9\,GHz. The synthesised beam is $10\arcsec \times 10\arcsec$ (PA $= 0\degr$). A 2D Gaussian fit yields a major axis of $10.07 \pm 0.03\arcsec$ and a minor axis of $9.96 \pm 0.03\arcsec$, aligned at $30\degr \pm 11\degr$. These values match the beam dimensions, indicating that H\,0557--385 is unresolved or marginally resolved, consistent with a compact morphology. No extended emission is detected around the source; the only large-scale feature in the field is a shallow negative bowl, a deconvolution artefact of the limited short-spacing coverage of the interferometric data (see Fig.~\ref{fig:fig1}).

\section{Broadband Radio Spectral Analysis}
\label{sec:jet_models}

To characterise the physical mechanism responsible for the compact radio emission in H\,0557--385 and to determine whether it is consistent with jet activity in a young radio galaxy or a more transient synchrotron process, we constructed a broadband radio spectrum spanning ${\sim}0.076$--258.8\,GHz by combining the newly acquired ATCA data with archival observations from MWA \citep{HurleyWalker2022, Ross2024}, ASKAP \citep{Duchesne2023, Duchesne2024, Duchesne2025}, and ALMA\footnote{ALMA project code 2016.1.01385.S} (see Table~\ref{tab:tab1}). The MWA data span 76--227\,MHz, the ASKAP data span 887\,MHz--1.655\,GHz, and the ALMA observations were taken at central frequencies of 240.842, 242.765, and 258.785\,GHz with 2\,GHz bandwidth.  Although the MWA, ATCA, ASKAP, and ALMA observations are not simultaneous and span 2017, 2020, 2021, 2022, and 2024, radio flux densities in the ASKAP bands are found to be stable within ${\sim}10$\%, a level of variability typical for Seyfert-type AGN and compact radio sources below the high-variability blazar regime \citep[e.g.,][]{Barvainis2005, Mundell2009, Baldi2015}.

The presence of a spectral turnover within the ATCA band (Section~\ref{sec:Imaging}) is confirmed when the MWA, ASKAP, and ALMA data extend the frequency coverage to 0.076--258.8\,GHz.  Such a turnover can arise from either SSA or FFA by surrounding thermal plasma \citep[see, e.g.,][and references therein]{Gregory1974, Miller-Jones2004}.

We evaluated both single-component alternatives before adopting the two-component model described below. A homogeneous external-screen FFA model \citep[e.g.,][]{Mezger1967, Bicknell1997, Callingham2015} was fitted to the full broadband spectrum and rejected: it predicts a steeply inverted flux density ($S_\nu \propto \nu^{-2.1}$ below the turnover) that falls more than three orders of magnitude below the observed ASKAP flux densities at ${\sim}0.9$--1.7\,GHz. A homogeneous SSA model \citep[e.g.,][]{Pacholczyk1970, Rybicki1979, Callingham2015} was similarly rejected: it requires an optically thin spectral index $\alpha_{\rm thin} = -1.76$, far steeper than observed in compact jets, and an optically thick slope $\alpha_{\rm thick} = +0.40$, far flatter than the canonical $+5/2$ for a uniform sphere.  Both failures point to a physically more complex, spatially stratified emission structure. We therefore attribute the ${\sim}9.4$\,GHz turnover of the compact component to synchrotron self-absorption in a stratified jet, rejecting both free-free absorption and homogeneous SSA for this component.

\subsection{Two-Component Stratified Synchrotron Model}
\label{sec:inh_ssa}

\begin{figure*}
    \includegraphics[width=2.0\columnwidth]{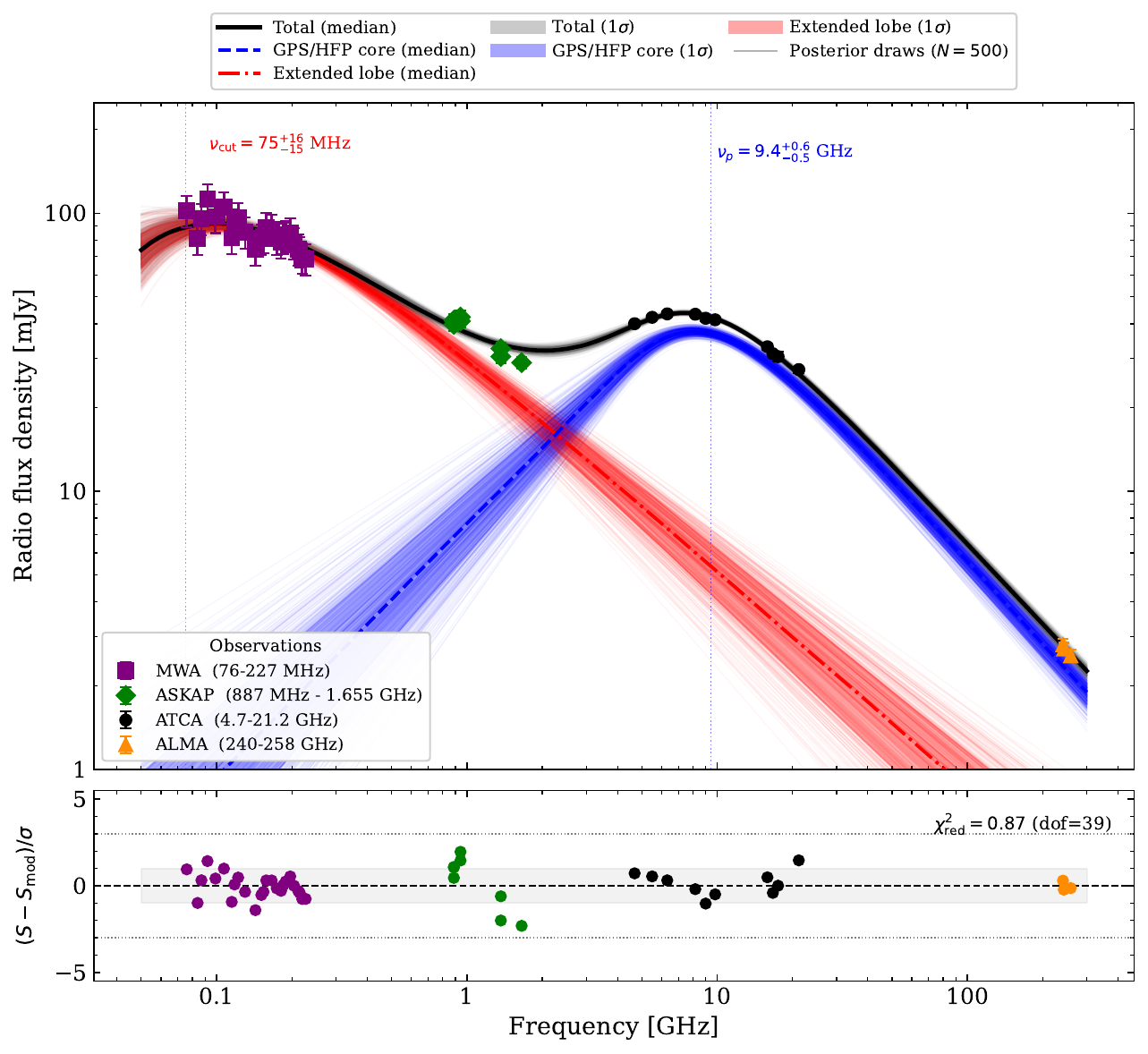}
    \caption{Broadband radio spectrum of H\,0557--385 [MWA: 76--227\,MHz \citep{HurleyWalker2022, Ross2024}; ASKAP: 887\,MHz--1.655\,GHz \citep{Duchesne2023, Duchesne2024, Duchesne2025}; ATCA: this work; ALMA: project code 2016.1.01385.S] with the best-fit two-component model comprising a stratified, self-absorbed component with a GPS/HFP spectral shape (blue dashed line) and an extended optically thin power-law lobe (red dash-dotted line). The solid black line shows the total median posterior; shaded bands show $1\sigma$ posterior uncertainty from 500 draws. Residuals normalised by $1\sigma$ errors are shown in the lower panel ($\chi^2_{\rm red} = 0.87$, 39\,dof). Systematic uncertainties of 10\% for MWA and 5\% for ASKAP, ATCA, and ALMA \citep{Partridge2016} have been added in quadrature. The compact self-absorbed component (turnover at $9.4^{+0.6}_{-0.5}$\,GHz) traces the compact self-absorbed jet base; the power-law lobe ($\alpha_{\rm pl}=-0.78^{+0.06}_{-0.07}$) traces extended, optically thin plasma that dominates the spectrum below ${\sim}1.6$\,GHz. Note: For MWA, in-addition to Table~\ref{tab:tab1} data points, we included extra inband data points for broadband spectral modelling.}
    \label{fig:fig2}
\end{figure*}

Inspection of the low-frequency MWA and ASKAP data reveals that this emission cannot be explained by the high-frequency GPS peak alone. The flux densities at ${\sim}0.08$--1.6\,GHz follow a simple declining power law that is physically distinct from the rising, self-absorbed GPS spectrum; the best-fitting compact self-absorbed component alone accounts for only ${\sim}7$\,mJy of the ${\sim}40$\,mJy observed at 887\,MHz. We therefore adopt a \textit{two-component} description comprising (i) a compact, self-absorbed component and (ii) an extended lobe described by a simple power law with an exponential cutoff (Fig.~\ref{fig:fig2}):
\begin{equation}
\label{eq:full_model}
S_{\rm model}(\nu) = S_{\rm GPS}(\nu) + S_{\rm PL}(\nu),
\end{equation}
where
\begin{align}
k &= \left(\frac{\nu}{\nu_{\rm p}}\right)\\
\label{eq:inh_model}
S_{\rm GPS}(\nu) &=\frac{S_{\rm p}}{\left(1-e^{-1}\right)} \left(k\right)^{\alpha_{\rm thick}}
        \left[1 - \exp\!\left(-\!\left(k
        \right)^{(\alpha_{\rm thin} - \alpha_{\rm thick})}\right)
        \right], \\
\label{eq:pl_model}
S_{\rm PL}(\nu) &= A_{\rm pl}\left(\frac{\nu}{\nu_{\rm ref}}\right)^{\alpha_{\rm pl}}\exp\!\left(-\frac{\nu_{\rm cut}}{\nu}\right),
\end{align}
with $\nu_{\rm ref} = 1$\,GHz. Here $S_{\rm p}$ is the GPS peak flux density at the turnover frequency $\nu_{\rm p}$; $\alpha_{\rm thin}$ and $\alpha_{\rm thick}$ are the optically thin (high-frequency) and optically thick (low-frequency) spectral indices respectively. Equation~\ref{eq:inh_model} is a generalisation of the standard peaked-spectrum function of \citet{Snellen1998}, popularised by \citet{Callingham2017}. $A_{\rm pl}$ and $\alpha_{\rm pl}$ are the normalisation and spectral index of the extended lobe component, and $\nu_{\rm cut}$ is an exponential low-frequency cutoff that suppresses the lobe emission below $\nu_{\rm cut}$ (see below). The total model has seven free parameters: $\{S_{\rm p},\,\nu_{\rm p},\,\alpha_{\rm thin},\,\alpha_{\rm thick},\,A_{\rm pl},\,\alpha_{\rm pl},\,\nu_{\rm cut}\}$.

Throughout this paper, the ``compact, self-absorbed component'' denotes the high-turnover component responsible for the gigahertz-peaked (GPS/HFP) spectral shape, as distinct from the extended, optically thin lobe. It traces the self-absorbed synchrotron plasma in the innermost parsecs of the jet. In stratified jet models, synchrotron emission at different frequencies arises from different sections along a conical jet; higher-frequency radiation emerges from the optically thick base near the black hole, while lower-frequency emission is produced further downstream \citep{Blandford1979, Konigl1981, Marscher1980}. The composite emission from these stratified regions produces the smooth spectral transition between the optically thick and thin regimes that characterises a peaked spectrum \citep{Marscher1980, Kaiser2006}.

The power-law component captures extended, optically thin lobe emission. Its presence as a distinct component is physically expected: as radio plasma expands and ages, the high-energy electrons responsible for self-absorbed emission lose energy, leaving a steep, non-absorbed synchrotron spectrum from the surrounding extended lobe \citep{Murgia1999, ODea1998}. H\,0557--385 is unresolved in all of our images, including the ALMA data, whose ${\approx}0.04\arcsec \times 0.03\arcsec$ beam corresponds to ${\approx}28 \times 21$\,pc at the source; the compact self-absorbed component and the extended lobe are therefore distinguished spectrally rather than spatially: below ${\sim}1.6$\,GHz the total flux density is dominated by the extended lobe, while above ${\sim}2$\,GHz the compact component dominates (Fig.~\ref{fig:fig2}).

We fitted the two-component model to the full 46-point broadband dataset [MWA: 25 points (including additional in-band measurements beyond those listed in Table~\ref{tab:tab1}); ASKAP: 8; ATCA: 10; ALMA: 3] using a Markov Chain Monte Carlo (MCMC) approach. The MWA points, including all in-band measurements used here, and their associated flux-density uncertainties were taken from the GLEAM-X catalogue \citep{HurleyWalker2022, Ross2024}; Table~\ref{tab:tab1} lists only a representative subset of these for compactness, whereas the fit uses the full set of GLEAM-X in-band measurements. The likelihood uses the $1\sigma$ flux-density uncertainty of each point (the catalogue error for MWA, and the quadrature sum of the Gaussian-fit and image-rms errors for ASKAP, ATCA, and ALMA), with an additional systematic uncertainty of 10 per cent for MWA and 5 per cent for ASKAP, ATCA, and ALMA \citep{Partridge2016} added in quadrature. Both components were evaluated simultaneously at all frequencies, such that the plotted total model represents the physical sum, $S_{\rm GPS}(\nu) + S_{\rm PL}(\nu)$, at each frequency. We adopted Gaussian priors for $S_{\rm p}$, $\nu_{\rm p}$, $\alpha_{\rm thin}$ and $\nu_{\rm cut}$, centred on the results of an initial fit to the same data, and uniform priors for the remaining parameters (Table~\ref{tab:fit_params}); in each case the posterior is substantially narrower than the prior, so the results are dominated by the data. The optically thick spectral index, $\alpha_{\rm thick}$, was allowed to vary over the range $0 \leq \alpha_{\rm thick} \leq 2.5$, extending from the relatively flat spectra expected for stratified synchrotron-emitting jets to the canonical homogeneous SSA limit of $\alpha_{\rm thick}=2.5$ \citep{Blandford1979,Konigl1981}. The MCMC analysis was performed using 64 walkers for 100,000 steps, with the first 25,000 steps discarded as burn-in. Convergence was assessed by requiring the length of the retained chain to exceed 50 times the integrated autocorrelation time for every fitted parameter.

Our best-fit parameters are $S_{\rm p} = 36.9^{+0.8}_{-0.9}$\,mJy, $\nu_{\rm p} = 9.4^{+0.6}_{-0.5}$\,GHz, $\alpha_{\rm thick} =+0.91\pm0.12$, $\alpha_{\rm thin} = -0.99\pm0.02$, $A_{\rm pl} = 31.8^{+1.9}_{-2.3}$\,mJy, $\alpha_{\rm pl} = -0.78^{+0.06}_{-0.07}$ and $\nu_{\rm cut} = 75^{+16}_{-15}$\,MHz, with $\chi^2_{\rm red} = 0.87$ (39 degrees of freedom). The model produces excellent agreement with the observational data across nearly four decades in frequency (Fig.~\ref{fig:fig2}); corner plots for all parameters are shown in Fig.~\ref{fig:figA1} of Appendix.

The optically thick slope ($\alpha_{\rm thick} = +0.91\pm0.12$) is substantially flatter than the canonical homogeneous SSA value of $+2.5$, yet steeper than expected for a simple stratified jet, pointing to a spatially structured absorber that combines synchrotron-emitting regions with a range of optical depths rather than a single uniform component \citep[e.g.,][]{Blandford1979, Konigl1981}. The optically thin index ($\alpha_{\rm thin} = -0.99\pm0.02$) implies an electron energy distribution $p = 2.98 \pm 0.04$ \citep[via $\alpha = -(p-1)/2$; e.g.,][]{Pacholczyk1970, Rybicki1979}, consistent with first-order Fermi (diffusive shock) acceleration at a strong shock associated with the jet \citep{Begelman1984}. We note that, because the source is spatially unresolved, our data constrain the presence of shock-accelerated electrons but not the location of the shock; we therefore do not associate it specifically with a jet termination point. The optically thin spectral index of the compact, self-absorbed component ($\alpha_{\rm thin} = -0.99\pm0.02$) and the spectral index of the extended lobe ($\alpha_{\rm pl} = -0.78^{+0.06}_{-0.07}$) are both consistent with steep-spectrum, optically thin synchrotron emission. At first sight the lobe index being flatter than that of the compact component may seem counter-intuitive, since relic plasma is expected to be more, not less, spectrally aged. This suggests that the lobe is unlikely to be simply aged plasma shed by the currently active component: passive ageing of the compact-component electrons ($\alpha_{\rm thin}=-0.99$) would steepen the spectrum further (towards $\alpha\lesssim-1.2$), whereas the lobe is observed to be flatter. Its index is instead close to a freshly injected value ($\alpha_{\rm inj}\approx-0.7$), which is more naturally explained if the lobe is a physically distinct electron population --- either injected in a separate, earlier episode, or continually replenished/re-accelerated in situ (for example at the interface where the outflow meets the ISM) --- rather than a passively ageing continuation of the present outflow. We return to the two evolutionary interpretations of this component in Section~\ref{sec:phys_interp}.

The low-frequency cutoff of the extended lobe is well constrained at $\nu_{\rm cut} = 75^{+16}_{-15}$\,MHz, placing it at the low-frequency edge of the MWA band, where the spectral flattening is directly observed: the MWA flux densities are nearly flat across 76--227\,MHz, whereas an un-cut power law with $\alpha_{\rm pl} = -0.78$ would continue to rise steeply toward lower frequencies and substantially overpredict the lowest-frequency points. The cutoff is therefore a genuine feature required by the GLEAM-X data and not a mathematical artefact: without it, the summed two-component model overpredicts the observed MWA flux densities.

To quantify this, we compared our adopted model against the simpler alternative of a compact self-absorbed component plus a plain power law with no low-frequency cutoff and the lobe spectral index left free, fitted to the identical dataset within the identical simultaneous framework (both components evaluated at all frequencies), with the same sampler, priors and error treatment. The cutoff model gives $\chi^2 = 34.0$ ($\chi^2_{\rm red} = 0.87$), while the no-cutoff model gives $\chi^2 = 61.4$ ($\chi^2_{\rm red} = 1.54$). The improvement of $\Delta\chi^2 = 27.4$ for one additional parameter is decisive on both the Akaike and Bayesian information criteria, which penalise the extra parameter yet still strongly favour the cutoff model ($\Delta{\rm AIC} = 25.4$, $\Delta{\rm BIC} = 23.5$, both in its favour; values $\gtrsim 10$ are conventionally regarded as decisive). Without the cutoff the plain power law cannot reproduce the flat MWA band and the fit is driven into a compromise that degrades both ends of the spectrum---the lobe index flattens to $\alpha_{\rm pl} = -0.55$ while $\alpha_{\rm thin}$ steepens to $-1.16$ and $\nu_{\rm p}$ rises to $10.9$\,GHz---yet it still overpredicts the lowest-frequency (76\,MHz) point by ${\sim}37$ per cent. The cutoff model requires none of these adjustments. We note that a plain-power-law lobe was in fact our starting model; the exponential cutoff was adopted only after we found it necessary to reproduce the observed low-frequency flattening.

We adopt a fixed exponent of unity in the cutoff term, providing the minimal one-parameter phenomenological description required to reproduce the observed low-frequency turnover. Although the cutoff frequency is reasonably constrained by the spectral flattening across the MWA band, the sharpness of the turnover is not. The cutoff lies near the low-frequency boundary of our observations, and the spectrum below it---where the detailed shape of the turnover would be most strongly imprinted---is only sparsely sampled. Allowing the exponent to vary freely would therefore introduce an additional parameter that is strongly correlated with $\nu_{\rm cut}$ and poorly constrained by the available data. A steeper frequency dependence, such as the $\nu^{-2.1}$ dependence expected for free-free absorption, would produce a sharper turnover, but the present data do not provide sufficient frequency coverage to distinguish between different cutoff shapes. We therefore treat the exponential cutoff purely as a phenomenological description of the observed spectrum rather than as evidence for a specific physical mechanism.

We now turn to the physical origin of the low-frequency suppression of the \emph{extended lobe} (the exponential cutoff at $\nu_{\rm cut}\approx75$\,MHz), which is a distinct question from the self-absorbed turnover of the compact component discussed above. One possible physical origin of this low-frequency cutoff is FFA by ionised gas along the line of sight to the radio-emitting region. Adopting the standard optical-depth relation $\tau_{\rm ff} \simeq 8.2\times10^{-2}\,T_{\rm e}^{-1.35}\,\nu_{\rm GHz}^{-2.1}\,{\rm EM}$ \citep{MezgerHenderson1967}, where the emission measure ${\rm EM} = \int n_{\rm e}^{2}\,\mathrm{d}l$ (in pc\,cm$^{-6}$) is the integral of the squared electron density along the path through the ionised screen, and $T_{\rm e}$ is the electron temperature, and requiring $\tau_{\rm ff}=1$ at $\nu_{\rm cut}$, gives an emission measure of ${\rm EM}\sim1.3\times10^{4}$\,pc\,cm$^{-6}$ for $T_{\rm e}=10^{4}$\,K. We adopt $T_{\rm e}=10^{4}$\,K as the canonical temperature of photoionised gas in AGN, characteristic of both the narrow-line region and the warm ionised ISM \citep{Osterbrock2006}; the inferred EM scales only weakly with this choice ($\propto T_{\rm e}^{1.35}$), changing by a factor of ${\lesssim}2.5$ (and $n_{\rm e}$ by ${\lesssim}1.6$) over the plausible range $T_{\rm e}\approx5\times10^{3}$--$2\times10^{4}$\,K. For a uniform absorbing screen with a characteristic path length of 1--100\,pc, this corresponds to electron densities of approximately $n_{\rm e}\approx12$--$115$\,cm$^{-3}$. Such densities are plausible for ionised gas in the nuclear regions of Seyfert galaxies and are intermediate between those typically associated with diffuse warm ionised gas and denser narrow-line-region clouds \citep[$n_{\rm e}\sim10^{2}$--$10^{4}$\,cm$^{-3}$ in the NLR;][]{Osterbrock2006, Peterson1997}. The required emission measure is also lower than the ${\rm EM}\sim10^{5}$--$10^{7}$\,pc\,cm$^{-6}$ values inferred for some GPS sources with GHz-frequency turnovers \citep{Bicknell1997,Callingham2015}, as expected for the substantially lower turnover frequency observed in H\,0557--385. Thus, an external FFA screen provides a physically viable explanation for the observed low-frequency suppression.

An alternative explanation is that this low-frequency cutoff of the extended lobe is intrinsic to the synchrotron-emitting plasma and arises from a low-energy cutoff in the relativistic electron population. Synchrotron radiation at a characteristic frequency $\nu$ is produced by electrons with Lorentz factor $\gamma\simeq(\nu/C_{B}B)^{1/2}$, such that a minimum Lorentz factor $\gamma_{\rm min}$ can introduce a corresponding low-frequency limit to the synchrotron emission. For the lobe magnetic-field strength derived in Section~\ref{sec:blobe} ($B_{\rm lobe}\approx0.5$--$3.8$\,mG), reproducing a characteristic cutoff near $\nu_{\rm cut}=75$\,MHz requires $\gamma_{\rm min}\approx70$--$190$, corresponding to electron energies of approximately $E_{\rm min}\approx35$--$95$\,MeV. These values are physically plausible for the low-energy end of a relativistic electron population, whose distribution may be determined by the particle-injection process and subsequently modified by Coulomb losses and other energy-dependent processes \citep{Pacholczyk1970,Longair2011}. The inferred values should nevertheless be regarded as order-of-magnitude estimates, since the precise relationship between the observed turnover frequency and $\gamma_{\rm min}$ depends on the shape of the electron energy distribution and the magnetic-field structure.

Independent support for the presence of circumnuclear ionised gas in H\,0557--385 comes from X-ray spectroscopy. \citet{Longinotti2009} and \citet{Coffey2014} find that the source harbours a persistent two-phase warm (ionised) absorber, detected in every spectral state, with column densities $N_{\rm H}\sim10^{21}$--$10^{22}$\,cm$^{-2}$ and ionisation parameters $\log\xi\sim0.3$--$2.2$, in addition to the variable neutral absorber responsible for its changing-look X-ray behaviour. \citet{Coffey2014} localise the circumnuclear absorbing structure to the dusty torus, estimating an inner (dust-sublimation) radius $R_{\rm d}\simeq0.6$\,pc and an outer edge $R_{\rm out}\simeq18$\,pc. This is directly relevant to our radio spectrum: the ionised gas required to produce free-free absorption at $\nu_{\rm cut}$ ($n_{\rm e}\approx12$--$115$\,cm$^{-3}$ over a 1--100\,pc path) is fully consistent, in both density and physical scale, with the circumnuclear ionised medium the X-ray data independently require. The X-ray observations therefore lend physical plausibility to the FFA interpretation of the low-frequency turnover, although they do not by themselves rule out an intrinsic electron cutoff.

Both FFA and an intrinsic low-energy cutoff therefore provide physically plausible explanations for the observed low-frequency suppression of the extended lobe. With the current frequency coverage, neither can be excluded outright, because the turnover is sampled over a limited frequency range and the phenomenological cutoff exponent is not independently constrained. On balance, however, we consider FFA the more likely of the two: the ionised-gas density and physical scale it requires ($n_{\rm e}\approx12$--$115$\,cm$^{-3}$ over 1--100\,pc) coincide with the clumpy, circumnuclear warm absorber that X-ray spectroscopy independently detects in this source \citep{Longinotti2009, Coffey2014}. We therefore adopt FFA as our preferred interpretation while retaining the intrinsic cutoff as a viable alternative. Observations extending to frequencies below ${\sim}70$\,MHz would provide substantially greater leverage on the turnover shape, while spatially resolved low-frequency spectroscopy and polarimetry could further test for signatures of an external absorbing screen. At the far-southern declination of H\,0557--385 ($\delta \approx -38\degr$), the relevant low-frequency facilities are the southern-hemisphere arrays: deeper and lower-frequency MWA observations (including the sub-70\,MHz band) and, in the future, SKA-Low, both located in Western Australia, will be important for determining the physical origin of the turnover and for clarifying the nature of the compact radio-emitting region in H\,0557--385.

\section{Physical Properties of H\,0557--385}
\label{sec:physical_params}

\subsection{Angular and Linear Size}
\label{sec:size}

H\,0557--385 is unresolved in all of our images. The tightest direct limit comes from the ALMA data, whose ${\approx}0.04\arcsec\times0.03\arcsec$ beam limits the millimetre emission to ${\lesssim}28$\,pc at the source (for comparison, the $10\arcsec$ ATCA beam corresponds to ${\sim}7.0$\,kpc). We therefore estimate the size of the compact component from its turnover frequency. We use the empirical correlation between the turnover frequency and the linear size established for GPS, CSS, and HFP sources \citep[e.g.,][]{Fanti1990, ODea1997, ODea1998, Keim2019}:
\begin{equation}
\label{turnover_size}
\log\,\nu_{\rm p} =
  -0.21\,(\pm0.05) - 0.65\,(\pm0.05)\,\log\,l_{\rm linear},
\end{equation}
where $\nu_{\rm p}$ is the turnover frequency in GHz and $l_{\rm linear}$ is the linear size in kpc. This gives a linear size of $15.0^{+6.3}_{-5.0}$\,pc (Fig.~\ref{fig:figA2}), corresponding to an angular size of $21.5^{+9.1}_{-7.2}$\,mas for an angular-diameter distance of 144.11\,Mpc (luminosity distance 154.03\,Mpc) \citep{Hogg1999, Wright2006, Planck2020, Planck2021}. This is consistent with the ${\lesssim}28$\,pc limit from the ALMA imaging. The turnover--size relation is calibrated on the overall extent of peaked-spectrum sources, so this estimate refers to the radio source as a whole. The region that is self-absorbed at $\nu_{\rm p}$ is much smaller: for a brightness temperature at the equipartition value of ${\sim}5\times10^{10}$\,K \citep{Readhead1994}, it would be only ${\sim}0.1$\,mas (${\sim}0.07$\,pc) across \citep{Condon1982}, resolvable only with VLBI.

\subsection{Equipartition Magnetic Field Strength}
\label{sec:bfield}

Since H\,0557--385 is spatially unresolved, we do not use the synchrotron self-absorption field estimator of \citet{Jones1974}, \citet{Kellermann1981}, and \citet{Marscher1983}, which scales as the fourth power of the angular size, so that even small size uncertainties lead to order-of-magnitude errors in the derived field strength. We instead estimate the equipartition magnetic field strength ($B_{\rm eq}$) from the minimum-energy formalism of \citet{Miley1980}, which depends only weakly on the assumed size ($B_{\rm eq}\propto l^{-6/7}$):
\begin{equation}
\label{eq:mag_field}
B_{\rm eq}\,[\mathrm{G}] =
  1.4 \times 10^{-4}\,(1+z)^{1.1}\,\nu_{\rm p}^{0.22}
  \left[\frac{S_{0}}{\theta_{\rm x}\,\theta_{\rm y}\,l_{\rm linear}}
  \right]^{2/7},
\end{equation}
where $z$ is the redshift, $\nu_{\rm p}$ is the spectral turnover frequency in GHz, $S_{0}$ is the radio flux density at $\nu_{\rm p}$ in Jy, $\theta_{\rm x}$ and $\theta_{\rm y}$ are the angular dimensions of the source in arcseconds (here $\theta_{\rm x} = \theta_{\rm y} = \theta$, the angular size from Section~\ref{sec:size}), and $l_{\rm linear}$ is the path length through the emitting region in kpc. Equation~\ref{eq:mag_field} assumes equal energy in relativistic protons and electrons ($k=1$), a unit filling factor, $\sin\phi = 1$, and an $\alpha = -0.8$ spectrum between 0.01 and 100\,GHz \citep{Miley1980}.

Using the source size derived in Section~\ref{sec:size}, we find $B_{\rm eq} = 2.8^{+1.2}_{-0.7}$\,mG (Fig.~\ref{fig:figA3}). This value is consistent with those commonly measured in young GPS and CSS sources \citep[e.g.,][]{Murgia1999, Orienti2014}. It is also comparable to the equipartition magnetic field strengths of $B_{\rm eq} \sim 2$\,mG derived for the parsec-scale nuclear regions of low-luminosity Seyfert galaxies by \citet{Bontempi2012} using EVN observations.

\subsubsection{Magnetic Field Strength of the Extended Lobe}
\label{sec:blobe}

The magnetic field strength derived above characterises the compact self-absorbed component; the extended power-law component traces a physically distinct plasma reservoir with its own field. Applying a minimum-energy calculation to the fitted power-law component (integrated over the rest-frame band 10\,MHz--100\,GHz, giving $L_{\rm syn} = 8.3\times10^{39}$\,erg\,s$^{-1}$, with uncertainties propagated from the MCMC posterior and $k=0$) gives $B_{\rm lobe} \simeq 0.5$--$3.8$\,mG for the size range derived below, with a fiducial value of $1.6$\,mG (for the turnover--size estimate of 15\,pc); adopting $k=1$, as for the compact component, would raise these values by a factor of $2^{2/7}\approx1.2$. The lobe is unresolved and its size is not constrained by our imaging, so we bound it by expansion physics: for expansion at $0.01$--$0.1c$, typical of GPS and CSS sources \citep{Polatidis2003, Orienti2014}, over the $\lesssim1.8^{+1.1}_{-0.7}\times10^3$\,yr synchrotron age it spans ${\sim}6$--$56$\,pc. Because $B_{\rm eq} \propto l^{-6/7}$, the factor of ${\sim}10$ range in size maps onto the factor of ${\sim}7$ range in $B_{\rm lobe}$ quoted above. The fiducial lobe field is thus lower than that of the compact self-absorbed component (2.8\,mG), as expected for a more diffuse plasma reservoir; we return to its significance in Section~\ref{sec:phys_interp}.

\subsection{Synchrotron Spectral Age}
\label{sec:synch_age}

We estimate the synchrotron spectral age ($t_{\rm spec}$) of the relativistic electron population following \citet{Miley1980}, which assumes continuous energy losses and no significant re-acceleration:
\begin{equation}
\label{eq:spec_age}
t_{\rm spec} = \frac{0.82\,\sqrt{B_{\rm eq}}}
                    {(B_{\rm eq}^2 + B_{\rm R}^{2})}
               \frac{1}{\sqrt{\nu_{\rm p}(1+z)}} \quad \text{[yr]},
\end{equation}
where $\nu_{\rm p}$ is the turnover frequency in GHz, and $B_{\rm R} = \sqrt{8\pi\,u_{\rm CMB}}\,(1+z)^{2} \approx 3.25\times10^{-6}\,(1+z)^{2}$\,G is the CMB-equivalent magnetic field \citep[from the radiation energy density $u_{\rm CMB}=4\sigma_{\rm SB}T_{\rm CMB}^4/c$;][]{Longair2011}. Equation~\ref{eq:spec_age} is the average age for an isotropic pitch-angle distribution and is defined for $\nu_{*}$, the frequency above which the spectrum steepens exponentially \citep{Miley1980}. No such steepening is detected up to 258.8\,GHz, so $\nu_{*} \gtrsim 259$\,GHz; adopting $\nu_{\rm p}$ instead therefore yields an upper limit on the radiative age. The $B_{\rm R}^{2}$ term represents inverse-Compton losses off the CMB. At the low redshift ($z=0.034$) and small size of H\,0557--385 the CMB is not expected to be the dominant inverse-Compton photon field; synchrotron self-Compton losses, or inverse-Compton scattering of the AGN and host-galaxy radiation fields, are likely to be more important \citep{Stawarz2008}. However, this does not affect our age estimate: the equipartition field derived here ($B_{\rm eq}\approx2.8$\,mG) exceeds $B_{\rm R}\approx3.5$\,\textmu G by nearly three orders of magnitude, so the source is deep in the synchrotron-loss-dominated regime ($B_{\rm eq}^{2}\gg B_{\rm R}^{2}$) and radiative cooling is overwhelmingly synchrotron. Any inverse-Compton contribution, whether from the CMB, the SSC photons, or galactic-scale radiation fields, would need an equivalent energy density comparable to $B_{\rm eq}^{2}/8\pi$ to matter, which is not the case here; the spectral age is therefore insensitive to the choice of inverse-Compton photon field.

Using the magnetic field derived in Section~\ref{sec:bfield}, we obtain a synchrotron age of $\lesssim(1.8^{+1.1}_{-0.7}) \times 10^3$\,yr for $B_{\rm eq} = 2.8^{+1.2}_{-0.7}$\,mG (Fig.~\ref{fig:figA4}). This upper limit is consistent with those of young radio sources such as GPS sources, suggesting that H\,0557--385 may host jets that are either still developing or have been re-ignited only recently.

\subsection{Jet Power Estimates}
\label{sec:jet_power}

We estimate the time-averaged jet power $Q$ under the equipartition assumption adopting spherical geometry, consistent with the compact morphology of the source. The source radius $R = l/2$ and volume $V = (4/3)\pi R^3$, where $l$ is the linear size, and the magnetic energy density is $u_{\rm B} = B_{\rm eq}^2 / (8\pi)$. In the equipartition limit $u_{\rm e} \approx u_{\rm B}$ \citep[e.g.,][]{Burbidge1956, Pacholczyk1970}.

We consider two physically motivated particle composition scenarios \citep[e.g.,][]{ODea1998, Orienti2014, Croston2018}:

\begin{enumerate}
    \item \textbf{Pair plasma (electrons + positrons)}: the total energy content is
    \begin{equation}
    E_{\rm pair} \approx 2\,u_{\rm B} V = \frac{B_{\rm eq}^2\,V}{4\pi}.
    \end{equation}

    \item \textbf{Proton-loaded plasma (one proton per radiating electron, with the same Lorentz factor)}: the proton energy density is
    \begin{equation}
    u_{\rm p} \approx \left(\frac{m_{\rm p}}{m_{\rm e}}\right) u_{\rm e},
    \end{equation}
    where $m_{\rm p}/m_{\rm e} \approx 1836.15$, giving $E_{\rm p} \approx (u_{\rm B} + u_{\rm e} + u_{\rm p})\,V$.
\end{enumerate}

The time-averaged jet power is $Q = E_{\rm total}/t_{\rm spec}$, where $t_{\rm spec}$ is the synchrotron spectral age. Uncertainties are propagated via the full MCMC posterior of $\nu_{\rm p}$ and $S_{\rm p}$ (240\,000 posterior draws), through which $B_{\rm eq}$, $l$, and $t_{\rm spec}$ are computed self-consistently on each draw. The pair-plasma case yields $Q_{\rm pair} = (5.5\pm0.4)\times10^{41}$\,erg\,s$^{-1}$ and the baryonic case $Q_{\rm proton} = 5.1^{+0.4}_{-0.3}\times10^{44}$\,erg\,s$^{-1}$. Notably, $Q$ does not depend on the assumed source size: in the synchrotron-dominated regime ($B_{\rm eq} \gg B_{\rm R}$), the minimum-energy $E_{\rm min} \propto V^{3/7}$ and the spectral age $t_{\rm spec} \propto V^{3/7}$ cancel exactly, so $Q = E_{\rm min}/t_{\rm spec}$ depends only on $\nu_{\rm p}$ and $L_\nu$, not on the assumed source size.

For H\,0557--385, $Q$ lies in the range ${\sim}5\times10^{41}$--$5\times10^{44}$\,erg\,s$^{-1}$ across the two plasma compositions (Table~\ref{tab:physical_comparison}), consistent with high-luminosity radio-quiet Seyferts and low-luminosity GPS/CSS sources of comparable size \citep[e.g.,][]{Morganti2003, KunertBajraszewska2010}.

We now compare this jet power with the accretion output of the AGN. From X-ray spectroscopy, \citet{Coffey2014} derive a black-hole mass of $M_{\rm BH}\sim6.4\times10^{7}\,M_{\odot}$ and a bolometric luminosity of $L_{\rm bol}\sim2.5\times10^{45}$\,erg\,s$^{-1}$, implying an Eddington luminosity $L_{\rm Edd}\sim8\times10^{45}$\,erg\,s$^{-1}$ and a (high-state) Eddington ratio $L_{\rm bol}/L_{\rm Edd}\sim0.3$. H\,0557--385 is therefore a radiatively efficient, sub-Eddington accretor, with an accretion mode typical of Seyfert\,1 nuclei rather than of the low-excitation, radiatively inefficient systems that host many powerful radio galaxies. The jet power represents a modest fraction of this output: $Q/L_{\rm bol}$ ranges from ${\sim}2\times10^{-4}$ (pair plasma) to ${\sim}0.2$ (proton-loaded), and $Q/L_{\rm Edd}$ from ${\sim}7\times10^{-5}$ to ${\sim}0.06$. Even in the proton-loaded case the jet carries at most ${\sim}20$ per cent of the bolometric luminosity, so it is energetically significant but not dominant over the radiative output. This combination---efficient, sub-Eddington accretion powering a jet that is present and non-negligible but sub-dominant---is consistent with a Seyfert-type nucleus that is currently launching, or has recently launched, a compact jet, and does not require the source to differ from ordinary Seyferts in its accretion properties. We note that, as a changing-look source, its bolometric luminosity and hence Eddington ratio vary between epochs \citep{Longinotti2009, Coffey2014}, so the value quoted here refers to the high state.

\begin{table}
\centering
\caption{Physical properties of H\,0557--385 derived from spectral modelling.  Jet power values are given for both the pair-plasma and proton-loaded assumptions (see text). The spectral age is an upper limit (Section~\ref{sec:synch_age}).}
\scalebox{1.0}{%
\begin{tabular}{lc}
\hline
\textbf{Parameter} & \textbf{H\,0557--385} \\
\hline
Turnover frequency (GHz) & $9.4^{+0.6}_{-0.5}$ \\
Linear size (turnover-relation, pc)           & $15.0^{+6.3}_{-5.0}$ \\
Angular size (turnover-relation, mas)         & $21.5^{+9.1}_{-7.2}$ \\
Magnetic field strength (turnover-size, mG)        & $2.8^{+1.2}_{-0.7}$ \\
Spectral age (turnover-size, yr)       &  $\lesssim(1.8^{+1.1}_{-0.7}) \times 10^3$ \\
\hline
Jet power (pair plasma, erg\,s$^{-1}$) & $(5.5\pm0.4)\times10^{41}$ \\
Jet power (proton-loaded, erg\,s$^{-1}$) & $5.1^{+0.4}_{-0.3}\times10^{44}$ \\
\hline
\end{tabular}}
\label{tab:physical_comparison}
\end{table}

\begin{table*}
\centering
\caption{Comparison of H\,0557--385 with other AGN classes. GPS/CSS/HFP source properties from \citet{ODea1998, Orienti2014, Dallacasa2000}; Seyfert properties from \citet{Laor2008, Panessa2019}; H\,0557--385 results from this work. Jet power values are given for the pair-plasma and proton-loaded assumptions (see Section~\ref{sec:jet_power}). Redshift and angular size are not compared, because their ranges largely reflect the selection of the comparison samples rather than intrinsic differences between the classes.  H\,0557--385 occupies an intermediate position: its spectral and size properties match young radio galaxies (GPS/HFP), while its energetics and host resemble radio-quiet Seyferts, making it a potential missing link between the two populations.}
\scalebox{0.73}{%
\begin{tabular}{l|c|ccc|l}
\toprule
\textbf{Property}
  & \textbf{H\,0557--385}
  & \textbf{GPS Sources\textsuperscript{a}}
  & \textbf{CSS Sources\textsuperscript{a}}
  & \textbf{HFP Sources\textsuperscript{c}}
  & \textbf{Radio-Quiet Seyferts\textsuperscript{b}} \\
\midrule
Peak Frequency (GHz) & $9.4$ & 0.5--10 & 0.1--0.5 & $>5$ & Often absent; \\
 &  &  &  &  & some flat-spectrum cores \\
Radio Power ($\log L_\nu$, W\,Hz$^{-1}$) & $23.0$ & $23.2$--$27.7$\textsuperscript{e} & $23.2$--$27.7$\textsuperscript{e} & 23--27 & 19--23\textsuperscript{f} \\
Radio Morphology & Compact (unresolved) & Compact ($\leq$1\,kpc) & Extended (1--20\,kpc) & Very compact ($<$100\,pc) & Diffuse/unresolved; \\
 &  &  &  &  & compact flat core in some \\
Linear Size (pc) & $15.0$ & 10--1000 & 1000--20,000 & $<$100 & $<$ few hundred \\
Spectral Shape & SSA turnover / smooth & Convex (SSA/FFA) & Power-law / curved & Sharp convex turnover & No peaked-spectrum component \\
 &  &  &  &  & observed; steep or flat-core \\
Spectral Age (yr) & $\lesssim1.8\times10^3$ & 100--10,000 & $10^5$--$10^6$ & $<$100 & -- \\
Jet Evidence & SSA + modelling & VLBI-scale jets & Core--lobe jets & VLBI cores with variability & Low-power, often subluminal jets \\
 &  &  &  &  & in a growing number of sources\textsuperscript{f} \\
Magnetic Field Strength (mG) & 2.8 & 1--100 & 0.1--10 & 10--100 & ${\sim}2$\textsuperscript{g}, $\ll$1 typical \\
Jet Power (erg\,s$^{-1}$) & $5.5 \times 10^{41}$ to $5.1\times10^{44}$\textsuperscript{d} & $10^{43}$--$10^{46}$ & $10^{43}$--$10^{46}$ & $10^{44}$--$10^{46}$ & $10^{38}$--$10^{42}$ \\
Host Galaxy Type & S0a (Seyfert\,1) & Elliptical / late-type & Elliptical & Often compact galaxies & Spiral \\
Star Formation Contribution & Low & Minimal & Sometimes strong & Likely negligible & Variable; not always dominant\textsuperscript{f} \\
\textbf{Classification} & \textbf{Intermediate / missing link} & Young radio galaxy & Young radio galaxy & Young radio galaxy & Radio-quiet AGN \\
\bottomrule
\end{tabular}}
\begin{flushleft}
\textsuperscript{a}\citet{ODea1998, Orienti2014};\quad
\textsuperscript{b}\citet{Laor2008, Panessa2019};\quad
\textsuperscript{c}\citet{Dallacasa2000};\quad
\textsuperscript{d} Values for pair plasma (lower bound) and proton-loaded (upper bound) assumptions. H\,0557--385 has the spectral morphology of a young radio galaxy but the energetics and host of a radio-quiet Seyfert.\quad
\textsuperscript{e}\citet{Nascimento2022};\quad
\textsuperscript{f}\citet{Panessa2019} and references therein report low-power, often subluminal jet signatures and a compact flat-spectrum radio core in a substantial fraction of radio-quiet AGN, with star-formation contributions spanning a wide range rather than uniformly dominating;\quad
\textsuperscript{g}\citet{Bontempi2012}.
\end{flushleft}
\label{tab:agn_comparison}
\end{table*}

\section{Discussion}
\label{sec:discussion}

\subsection{H\,0557--385 as a Candidate Missing Link Between Radio-Quiet and Radio-Loud AGN}
\label{sec:missing_link}

\subsubsection{Where does H\,0557--385 fit in the AGN landscape?}
\label{sec:where_fit}

The physical parameters derived in Section~\ref{sec:physical_params} place H\,0557--385 in an unusual position. Optically it has long been treated as radio-quiet, on the basis of its Seyfert\,1 spectrum and its radio-to-optical flux ratio \citep{Kellermann1989}. Yet every radio-physical quantity we can measure --- the spectral shape, the turnover frequency, the source compactness, the synchrotron age, and the equipartition magnetic field --- points instead to a young radio galaxy in the GPS or HFP phase. Table~\ref{tab:agn_comparison} sets these properties side by side and makes the tension explicit.

The turnover at ${\sim}9.4$\,GHz, the parsec-scale source size (${\sim}15$\,pc), and the synchrotron age of $\lesssim1.8\times10^3$\,yr are the classic signatures of newly ignited or recently re-triggered activity in compact young radio galaxies \citep{ODea1998, Orienti2014, Dallacasa2000}. In GPS and HFP sources these features arise because the expanding synchrotron plasma is still confined within or close to the nucleus, giving a self-absorbed peak that drifts to lower frequency as the source grows and ages \citep{Orienti2016, Callingham2017, Nyland2020}. H\,0557--385 shows such a self-absorbed peak; whether it drifts as predicted remains to be tested (Section~\ref{sec:two_scenarios}).

This placement is reinforced by the radio power--linear size ($P$--$D$) diagram, the standard diagnostic for the evolutionary state of radio sources \citep[e.g.,][]{ODea1997, An2012, Amenta2026}. With a linear size of ${\sim}15$\,pc and a monochromatic radio power of $\log(L_\nu/{\rm W\,Hz^{-1}})\approx23$, H\,0557--385 lies in the compact, low-power corner of the $P$--$D$ plane, coincident with the youngest and smallest GPS/HFP sources but offset to lower power than the classical radio-loud population \citep{Amenta2026}. Its position is consistent with an object at the earliest evolutionary stage, whether it subsequently grows along an evolutionary track (Scenario~A) or fades (Scenario~B).

It is also worth situating H\,0557--385 against the broader population of low-luminosity AGN (LLAGN). The LeMMINGs survey has shown that nuclear radio emission is common in LLAGN and low-power Seyferts, and that a substantial fraction hosts compact, often flat- or peaked-spectrum radio cores powered by low-level jet activity rather than star formation \citep{Baldi2021}. H\,0557--385 can be viewed as an extreme member of this continuum: its radio core is not merely compact but shows a well-defined, high-frequency self-absorbed peak, placing it at the young-radio-galaxy end of the range of jet activity seen in the LLAGN/Seyfert population, and reinforcing the view that the radio-quiet label conceals a diversity of underlying jet states.

At the same time, several properties associate the source with lower-power Seyfert systems rather than with classical radio galaxies. The inferred jet powers (${\sim}5\times10^{41}$--$5\times10^{44}$\,erg\,s$^{-1}$) reach only the lower end of those of classical radio-loud young radio galaxies ($10^{43}$--$10^{46}$\,erg\,s$^{-1}$). The host is an S0a (early-type disc) galaxy \citep{Ashton2005} rather than the massive elliptical that usually harbours a powerful GPS or CSS source, and the total radio luminosity is that of a Seyfert.

The radio properties themselves are what set H\,0557--385 apart. Radio-quiet Seyferts display a range of spectral shapes --- steep and declining, flat, or compact-core-dominated --- produced by star formation, coronal emission, or weak unconfined outflows \citep{Laor2008, Panessa2019}, but they do not show the well-defined self-absorbed peak seen here. Judged on spectral shape, turnover frequency, compactness, synchrotron age, and magnetic field strength, H\,0557--385 is instead consistent with young radio galaxies, while its total radio power remains at Seyfert levels. We therefore regard it as an intermediate object: a source with the radio characteristics of a young radio galaxy but the luminosity and host of a Seyfert.

The broader optical properties of peaked-spectrum sources support this reading. \citet{Nascimento2022} examined the optical spectra and hosts of 58 CSS/GPS and 14 MPS sources spanning $\log L_{\rm R} \sim 23.2$--$27.7$\,W\,Hz$^{-1}$, and found no strong link between the optical and radio properties: many are hosted by galaxies with intermediate-to-old stellar populations rather than being actively star forming. The GPS/HFP phenomenon is therefore not tied to any single optical signature, and a compact, self-absorbed radio source can reside in a host whose optical spectrum provides little indication of its radio properties. With a peak spectral luminosity of $\log L_\nu \approx 23.0$\,W\,Hz$^{-1}$, at or just below the low end of the \citet{Nascimento2022} range, H\,0557--385 is consistent with, rather than anomalous within, the known optical diversity of GPS-class sources.

\subsubsection{The radio-loud/radio-quiet classification is definition-dependent}
\label{sec:youngRG}

Beyond the sensitivities of the flux-ratio metric already noted, the deeper problem is that there is no single agreed definition to appeal to. The classical \citet{Kellermann1989} criterion uses the flux ratio $R = S_{\rm 5\,GHz}/S_{\rm 4400\,\angstrom} > 10$ for radio-loudness, whereas \citet{Kellermann2016}, working from a volume-limited sample of low-redshift QSOs at 6\,GHz, favour a spectral-luminosity threshold of $\log(L_{\rm 6\,GHz}/{\rm W\,Hz^{-1}}) \gtrsim 23.2$. \citet{Ho2001} take yet another route for low-luminosity AGN, noting that the radio-to-optical flux ratio breaks down once host starlight dilutes the nuclear emission. The choice of definition matters directly here: under the \citet{Kellermann2016} luminosity criterion, H\,0557--385 at $\log(L_{\rm 6\,GHz}) \approx 23.1$\,W\,Hz$^{-1}$ lies at the dividing line, so the classification is ambiguous by definition. For peaked sources the classification can also be biased by the standard 5\,GHz classifying frequency, which falls on the optically thick side of a spectrum peaking at ${\sim}9.4$\,GHz; for H\,0557--385 this effect is small (${\sim}3$ per cent), because the extended lobe, which contributes ${\sim}20$ per cent of the flux density at 5\,GHz, compensates for the optically thick decline of the compact component.

The radio-loudness classification of an AGN can also evolve over time, as the jet develops \citep{Nyland2020} or as the radio emission varies \citep{Bruni2026}; consequently, a classification based on a single-epoch observation may not reflect the source's long-term radio state. Taken together, these points show that H\,0557--385 cannot be placed cleanly on either side of the divide, and they motivate a physically grounded classification of the kind proposed by \citet{An2026}.

The synchrotron age of $\lesssim1.8\times10^3$\,yr is central here, because it constrains when the current episode of radio activity began, not the total age of the AGN or its accretion history. Young radio galaxies are defined by radio emission that is young, compact, and self-absorbed, whatever the properties of the black hole or disc \citep{ODea1998, An2012}. By that physical definition H\,0557--385 qualifies. The optical Seyfert classification does not contradict this: several confirmed GPS/CSS/HFP sources have Seyfert-like optical spectra, especially at lower redshifts and luminosities \citep{Giroletti2009, Congiu2017, Inoue2018}, and the combination of broad emission lines with a compact radio source is consistent with a scenario in which jet launching has only recently begun, perhaps triggered by a burst of accretion or magnetic flux accumulation near the black hole \citep{Narayan2003, Tchekhovskoy2011, Sikora2013}.

This view has recently gained direct observational support. VLBI monitoring of the historically radio-quiet narrow-line Seyfert\,1 Mrk\,110 revealed an intermittent, superluminal relativistic jet \citep{Wang2025}, with jet formation proceeding as a continuum of behaviours rather than a binary outcome within the radio-loud/radio-quiet framework \citep{An2026}. \citet{An2026} argue that a dichotomy defined by a single flux-ratio threshold cannot capture this continuum, and propose classifying AGN instead by the physical state of their jet activity --- persistent, intermittent, or apparently absent --- independent of the legacy flux-ratio boundary. H\,0557--385 is consistent with this framework: its compact, self-absorbed component (with a GPS/HFP-type spectrum) marks a currently active, intermittent or recently re-triggered jet phase, much like the one seen in Mrk\,110, regardless of its conventional optical label.

The same breakdown of the dichotomy is now seen at the high-power end. High-resolution studies of radio-quiet quasars find that a large fraction host compact, sometimes jet-like, radio structures whose emission is dominated by the AGN rather than by star formation, so that ``radio-quiet'' does not imply ``jet-free'' even for luminous quasars \citep[e.g.,][]{Jarvis2019, Njeri2026}. H\,0557--385 is the lower-luminosity, earlier-evolutionary-stage analogue of this behaviour: across four orders of magnitude in AGN power, from Seyferts to quasars, the radio-quiet population is increasingly found to contain genuine, if modest or intermittent, jet activity that the flux-ratio classification does not capture.

The practical consequence is that census estimates of the radio-loud fraction may undercount young GPS/HFP sources, especially at low luminosity and in disc-dominated hosts, precisely because the biases above push such sources below the radio-loud threshold. Broadband radio surveys with MeerKAT, ASKAP, ngVLA, and eventually the SKA --- which cover hundreds of MHz to tens of GHz in a single pass --- will be needed to find this population, and H\,0557--385 motivates targeted searches for GHz-peaked spectra within optically classified radio-quiet samples. High-frequency coverage above 10\,GHz is especially important: standard low-frequency surveys (e.g. NVSS, SUMSS, TGSS, RACS) would detect H\,0557--385 but would miss its turnover entirely, and only the combination of ATCA, ASKAP, and ALMA makes its nature as a compact, self-absorbed source unambiguous.

\subsubsection{Two interpretations and a path to discriminate them}
\label{sec:two_scenarios}

Two physically distinct scenarios are consistent with the present data.

\textit{Scenario A --- a young radio galaxy.} H\,0557--385 is in the early stages of jet development, analogous to a low-power GPS or HFP source. The compact, self-absorbed emission traces a newly launched or recently re-triggered jet whose plasma is still confined to the inner parsecs. If the fuelling is sustained, the jet may expand into a CSS source within ${\sim}10^{5}$\,yr and eventually, over $10^{7}$--$10^{8}$\,yr, into a large-scale radio galaxy \citep{ODea1998, KunertBajraszewska2010, An2020}. Here the radio-quiet label is simply an inadequate metric applied to a source at a critical evolutionary stage.

\textit{Scenario B --- a transient or episodic source.} H\,0557--385 is a Seyfert undergoing a brief episode of enhanced radio activity, perhaps driven by intermittent magnetic flux accumulation or a temporary magnetically arrested disc state \citep{Narayan2003, Czerny2009, Sikora2013}. In this case the phase of enhanced radio activity is short-lived, and the source will return to a radio-quiet state once the episode ends. Repeated episodes of this kind could, in aggregate, produce the ``radio-quiet'' AGN with elevated radio emission seen in population studies.

The two scenarios also map onto distinct jet-launching modes. A sustained young jet (Scenario~A) is most naturally powered by a magnetically arrested disc (MAD) state \citep{Narayan2003, Tchekhovskoy2011, Narayan2022}, in which spin energy extracted through the Blandford--Znajek mechanism \citep{Blandford1977} drives a compact, self-collimated jet; even at the milliGauss fields inferred here, MAD-like conditions near the innermost stable circular orbit can amplify the field enough to launch a jet for modest black-hole spin \citep{Tchekhovskoy2010}. An episodic source (Scenario~B) is better matched to magneto-centrifugal launching in a conventionally magnetised disc \citep{Blandford1982, Ferreira1997, Cao2014}, whose less efficient, less collimated outflows fit the relatively low jet power of H\,0557--385 and are punctuated by intermittent accretion or flux-transport events \citep{Czerny2009, Sikora2013, Dexter2014}.

Both place H\,0557--385 at the interface between the two classes, and both can be tested. The discriminators are VLBI-resolved jet morphology and proper motions, which would confirm collimated, advancing plasma in Scenario~A; multi-epoch tracking of the turnover frequency, which should drift steadily downward as the source expands in Scenario~A but vary erratically in Scenario~B; linear polarisation and Faraday rotation, which probe the ordered field of a sustained jet; and a check for core variability inconsistent with passive spectral evolution, the criterion \citet{Orienti2021} use to separate genuinely young sources from short-lived sources whose peaked spectra come from a variable core. Two caveats apply to these diagnostics. First, genuine young radio sources are typically weakly polarised, with strong internal and external Faraday depolarisation in the dense circumnuclear medium; high fractional polarisation would more likely signal a beamed, blazar-like contaminant than a young source \citep[e.g.,][]{ODea1998, Orienti2016}. Polarimetry is therefore most useful here as a check against beaming, rather than as a positive probe of an ordered jet field. Second, cleanly distinguishing the MAD and magneto-centrifugal (``CAD'') launching modes is observationally demanding: it requires resolving the jet base on sub-parsec scales, which for H\,0557--385 ($1\,{\rm mas}\approx0.7$\,pc) corresponds to milliarcsecond or finer resolution. This is beyond current connected-element arrays but is in principle within reach of high-frequency VLBI and, in the longer term, space VLBI; even then, multi-epoch tracking of the turnover is likely to remain the more practical near-term test of the two scenarios. Whichever scenario holds, H\,0557--385 shows that the optical radio-loud/radio-quiet boundary does not cleanly separate physically distinct populations, and that sources at this interface are well suited to studying how radio AGN activity is triggered, sustained, and switched off.

The X-ray data provide independent context for the accretion state that underlies these scenarios. The high-state Eddington ratio $L_{\rm bol}/L_{\rm Edd}\sim0.3$ (Section~\ref{sec:jet_power}; \citealt{Coffey2014}) shows that H\,0557--385 accretes efficiently in a standard, radiatively dominated Seyfert\,1 mode, rather than in the radiatively inefficient regime often associated with sustained, powerful radio jets. Efficient accretion of this kind is compatible with either scenario, but the pronounced X-ray variability of the source---its transitions between unabsorbed and Compton-thin absorbed states driven by clumpy circumnuclear gas \citep{Longinotti2009, Coffey2014}---indicates a dynamic nuclear environment of the kind in which intermittent fuelling or flux-transport events (Scenario~B) could plausibly occur.

It is also possible to specify the timescales over which the two scenarios can be discriminated. Under Scenario~A the turnover frequency should drift downward as the source expands: combining the size--turnover relation ($\nu_{\rm p}\propto l^{-0.65}$) with an expansion speed of $0.01$--$0.1c$ gives a fractional drift $\mathrm{d}\ln\nu_{\rm p}/\mathrm{d}t\sim0.65\,(v_{\rm exp}/l)$, i.e. ${\sim}0.01$--$0.2$ per cent per year for the overall size given by the turnover--size relation ($10$--$21$\,pc; Section~\ref{sec:size}) and these speeds. A detectable ${\sim}10$ per cent shift in $\nu_{\rm p}$ therefore accumulates over ${\sim}50$\,yr for the smallest, fastest-expanding case, but over ${\sim}10^{3}$\,yr for the largest, slowest case. A steady, monotonic drift measured over a multi-decade baseline would thus favour Scenario~A, whereas erratic or absent turnover evolution on the same baseline, or variability inconsistent with smooth expansion, would favour the episodic Scenario~B \citep[cf.][]{Orienti2021}. The synchrotron age of the source as a whole, $\lesssim1.8\times10^{3}$\,yr, also sets the expected duration of a single episode in Scenario~B, so a source that persists in its current state for much longer than this, or fades on a comparable timescale, would help distinguish the two pictures.

\subsection{Physical interpretation: jet activity in a young radio galaxy?}
\label{sec:phys_interp}

The two-component stratified synchrotron model (Section~\ref{sec:inh_ssa}; Fig.~\ref{fig:fig2}) gives a physically self-consistent description of the radio emission of H\,0557--385, with the two components tracing distinct plasma populations.

\textit{Compact self-absorbed component --- the currently active jet.} The compact self-absorbed component has an optically thick slope ($\alpha_{\rm thick} = +0.91\pm0.12$) that is flatter than the homogeneous SSA value of $+2.5$, together with a smoothly curved spectrum --- the signatures of a spatially stratified, partially self-absorbed conical jet \citep{Blandford1979, Marscher1980, Konigl1981} that spans a continuous range of optical depths along its axis: the highest-frequency emission arises from optically thick plasma near the black hole, while lower frequencies trace progressively more diffuse plasma downstream. The optically thin index ($\alpha_{\rm thin} = -0.99\pm0.02$; $p = 2.98\pm0.04$) is consistent with first-order Fermi acceleration at a strong shock in a newly launched jet \citep{Begelman1984}, and the failure of both homogeneous SSA and FFA models (Section~\ref{sec:jet_models}) confirms that a uniform emitting region cannot reproduce the observed curvature.

\textit{Extended power-law lobe.} The extended power-law component ($A_{\rm pl} = 31.8^{+1.9}_{-2.3}$\,mJy, $\alpha_{\rm pl} = -0.78^{+0.06}_{-0.07}$, with $\nu_{\rm cut} = 75^{+16}_{-15}$\,MHz), provides an independent constraint. Its index sits close to a freshly injected synchrotron spectrum ($\alpha_{\rm inj} \sim -0.7$), marking a distinct plasma population whose electrons are not simply an aged continuation of the compact component. Its minimum-energy magnetic field strength, $B_{\rm lobe} \simeq 0.5$--$3.8$\,mG (fiducial 1.6\,mG; Section~\ref{sec:blobe}), is on average weaker than that of the compact component, as expected for a more diffuse reservoir that has expanded away from the launching region. The coexistence of the young, self-absorbed compact component and the surrounding extended lobe gives a two-component view of the history of the source: the compact component marks the current phase of compact jet launching, while the lobe traces synchrotron plasma distributed more widely through the nuclear ISM. Under Scenario~A this is the growing, optically thin lobe of an expanding young radio galaxy; under Scenario~B it is plasma from a separate episode, implying that the present jet is not the first and that recurrent activity has progressively filled the nuclear ISM with synchrotron-emitting electrons. Either way, the lobe shows that synchrotron-emitting plasma is not confined to the compact self-absorbed component. A related two-component structure appears in the kiloparsec-scale radio knots recently found in the nearby low-luminosity AGN NGC\,7213 \citep{Bruni2026}, where flat-spectrum, high equipartition-field knots were read as compact termination shocks of a confined, recurrently active jet rather than classical lobes --- direct evidence that Seyfert-like AGN can retain a multi-epoch record of past jet activity like the one proposed here.

Consistent with the comparison presented in Section~\ref{sec:where_fit}, the inferred low jet powers place H\,0557--385 towards the low-power end of the young radio galaxy population, a regime that may be under-represented in observational samples owing to the limited sensitivity of existing radio surveys.

On galactic scales, jet--ISM interaction is usually identified morphologically, through distorted or asymmetric lobes, sharp bends, or dynamically disturbed gas revealed by optical/near-infrared kinematics \citep[e.g.,][]{Morganti2017, Jarvis2019}. For H\,0557--385 no such morphological evidence is available: the source is unresolved even in the ALMA images (${\approx}28$\,pc resolution), so we cannot search for disturbed lobe structure or measure the kinematics of the surrounding gas directly. The evidence for interaction is therefore indirect. If the low-frequency cutoff of the extended lobe is due to free-free absorption (Section~\ref{sec:jet_models}), it requires the lobe to be embedded in, or seen through, ionised gas ($n_{\rm e}\approx12$--$115$\,cm$^{-3}$); and the X-ray spectroscopy independently reveals a clumpy, variable circumnuclear absorber at parsec-to-tens-of-parsec scales \citep{Longinotti2009, Coffey2014}. Together these indicate that the jet is expanding into a dense and dynamic nuclear medium, precisely the environment in which jet--ISM interaction is expected, even though the interaction cannot yet be imaged. VLBI, together with spatially resolved optical/IR spectroscopy of the nuclear gas, would be needed to establish it morphologically and kinematically.

\subsection{Evolutionary implications: from Seyfert to radio galaxy?}
\label{sec:evolution}

The two scenarios of Section~\ref{sec:two_scenarios} imply very different futures. Under Scenario~A, sustained fuelling would let the plasma expand at a fraction of the speed of light, the turnover migrate to lower frequency along the linear-size--turnover relation (Equation~\ref{turnover_size}), and the source pass through the GPS, CSS, and eventually large-scale radio galaxy phases over $10^{7}$--$10^{8}$\,yr \citep{ODea1998, KunertBajraszewska2010, An2020}; H\,0557--385 would then represent the youngest, most compact stage of that sequence in an S0a (early-type disc) host \citep{Ashton2005}, a configuration allowed in principle but rarely seen. Under Scenario~B, the current episode fades on a timescale comparable to the synchrotron age and the source reverts to radio-quiet until the next episode; recurring ${\sim}10^3$\,yr episodes with a duty cycle of a few per cent would appear in population studies as a rare but not exceptional class of radio-variable Seyferts \citep{Saikia2009, Bruni2019, Brienza2020}.

Scenario~A must also be assessed against a well-established observational constraint: evolved, large-scale radio galaxies are almost exclusively hosted by massive elliptical galaxies, and are very rarely found in disc-dominated hosts \citep[e.g.,][]{Best2005, Tadhunter2016}. The host of H\,0557--385 is an S0a galaxy, which, while closer to the early-type regime than a pure spiral, is still a disc galaxy rather than the massive elliptical typical of classical radio galaxies. If H\,0557--385 were the progenitor of a source that will grow to large radio-galaxy scales, its early-type disc host would therefore make it an atypical, though not unprecedented, case, since the extended, evolved end-states of that evolutionary track are rarely observed in such hosts. This tension does not rule out Scenario~A --- the source could, for example, stall or fade before reaching large scales --- but it weighs somewhat against a straightforward, monotonic evolution into a classical radio galaxy, and correspondingly lends some support to Scenario~B, in which the radio activity is intrinsically short-lived and episodic and the source never evolves beyond the compact phase. The scarcity of evolved radio sources in disc-dominated hosts is at least as naturally reconciled with H\,0557--385 being a transient or recurrent compact source as with its being the youngest stage of a long-lived radio galaxy.

In either case, the jet power --- even at its lower bound of ${\sim}5\times10^{41}$\,erg\,s$^{-1}$ --- is enough to drive mechanical feedback into the circumnuclear ISM on sub-kiloparsec scales \citep{Morganti2017, Harrison2018, Mukherjee2018}. Even weak, confined jets can drive turbulence, ionise gas, and modulate star formation in the inner host \citep{Wagner2012, Morganti2015, Gaspari2017} --- effects not captured by radio-quiet AGN feedback models, but potentially important for the subset of Seyferts that pass through young-jet phases.

\subsection{Star formation as an alternative scenario}
\label{sec:sf}

Before attributing the radio emission to a jet, we consider whether star formation could account for it. The morphological, spectral, and host-galaxy evidence argues that it cannot.

The source is unresolved in all of our images, from the $10\arcsec$ ATCA beam (${\sim}7$\,kpc), in which no extended emission is detected, to the ${\approx}0.04\arcsec$ ALMA beam (${\approx}28$\,pc): it is a compact, core-like object with none of the extended features expected from star formation, such as kiloparsec-scale synchrotron haloes or clumpy thermal emission from H\,{\sc ii} regions \citep[e.g.,][]{Roy1994, Colbert1996, Thean2001, Gallimore2006}. The spectral shape is also inconsistent with star formation: it shows a smooth GPS/HFP turnover at ${\sim}9.4$\,GHz that requires a two-component stratified synchrotron model (Section~\ref{sec:inh_ssa}), whereas star-forming galaxies follow a single power law without such a turnover \citep[e.g.,][]{Klein2018, Galvin2018, VLA-COSMOS2019}.

The energetics lead to the same conclusion. Using archival mid-infrared photometry, the compact nuclear MIR emission in the sub-arcsecond atlas of \citet{Asmus2014}, and available \textit{IRAS}/WISE measurements, we estimate a conservative upper limit on the star-formation rate of ${\rm SFR} \lesssim 1$--$3\,M_{\odot}\,{\rm yr}^{-1}$, consistent with the \textit{Herschel}-based results of \citet{Shimizu2017}. With standard radio--SFR calibrations \citep{Yun2001, Murphy2011}, this corresponds to $L_{\rm 1.4\,GHz} \lesssim (1.6$--$5.1)\times10^{21}\,{\rm W\,Hz}^{-1}$, and at $D_{\rm L}=154.03\,{\rm Mpc}$ to an expected flux of only $S_{\rm 1.4\,GHz} \lesssim 0.6$--$1.8\,{\rm mJy}$ --- more than an order of magnitude below the observed ${\sim}31$--$33$\,mJy at 1.37\,GHz (Table~\ref{tab:tab1}). Even allowing for the $\sim0.3$\,dex scatter in the far-infrared--radio correlation \citep{Yun2001, Murphy2011}, star formation cannot supply the observed radio emission.

The infrared and host properties support the same conclusion. Mid-IR diagnostics from archival WISE and Spitzer data show emission dominated by hot dust heated by the AGN torus, with minimal PAH contribution, and the W3--W4 colours and IRAC flux ratios fall in the AGN-dominated regions of standard diagnostic diagrams \citep[e.g.,][]{Stern2012, Assef2013}. The presence of a substantial dusty torus is consistent with the circumnuclear obscuring structure inferred from the X-ray absorption \citep[$0.6\lesssim R\lesssim18$\,pc;][]{Coffey2014} and with recent detailed models of the clumpy, dusty tori of nearby AGN \citep{Drewes2025}; the same dense, structured circumnuclear medium provides the natural reservoir of ionised gas invoked in the free-free-absorption interpretation of the low-frequency turnover (Section~\ref{sec:jet_models}). The host is an S0a galaxy \citep{Ashton2005}, a class that typically shows modest or quenched star formation \citep{Kormendy2004, Cappellari2011}, its optical spectrum is that of a Seyfert\,1, with no strong starburst signatures in standard emission-line diagnostics \citep{Kewley2006}, and archival IR photometry (IRAS/AKARI/Herschel) shows no far-infrared excess of the kind seen in starburst galaxies \citep[cf.][]{Sanders1996}. The compact morphology, the peaked spectrum, the far-infrared--radio correlation, and the quiescent host together rule out star formation as the origin of the radio emission and reinforce the young-jet interpretation.

\section{Caveats and Limitations}
\label{sec:caveats}

While our multi-frequency analysis indicates that H\,0557--385 is a compact, self-absorbed synchrotron source in the GPS/HFP regime---and is therefore better classified as an intermediate or young-radio-galaxy-like object than as a canonical radio-quiet AGN---several limitations should be noted when interpreting the specific physical parameters and evolutionary scenarios.

\begin{itemize}

\item \textbf{Non-simultaneity of observations:} The radio spectra are compiled from non-simultaneous observations spanning MWA (2020), ASKAP (2020--2024), ATCA (2022), and ALMA (2017). Radio variability is common in Seyfert-type AGN but is generally of low amplitude. \citet{Hernandez-Garcia2026} detect significant variability in 86 per cent of a hard X-ray selected sample, yet find a median fractional rms amplitude of only 10 per cent (4--23 per cent) in the radio band, well below the corresponding optical and X-ray values. \citet{Mundell2009} likewise report no detectable variability in half of their sample, with the largest amplitude reaching ${\sim}38$ per cent, and \citet{Bruni2026} measure fractional variability amplitudes of $5.8\pm1.1$ and $10.0\pm1.0$ per cent at 5.5 and 9.0\,GHz respectively, the source being more variable at the higher frequency; individual objects may show no core variability at all \citep{Jones2011}. Variability at the ${\sim}10$ per cent level is thus expected, which is small compared with the spectral features modelled here, but intrinsic variability or secular evolution in the jet structure could still affect our results. Simultaneous broadband observations are necessary to better constrain spectral turnover features and potential variability signatures. More generally, sparse frequency coverage and the effects of spatial and temporal filtering can introduce ambiguities in determining the underlying absorption physics in spectral modelling of compact radio AGN \citep{Patil2022}; our broadband 0.076--258.8\,GHz dataset substantially mitigates but does not eliminate these concerns.

\item \textbf{Assumed source geometry and homogeneity:} Our two-component model captures the spectral structure of the compact self-absorbed component and the extended lobe but is based on spatially unresolved data.  It does not account for complex sub-structure within each component (e.g., multiple knots, jet bends, or lobe asymmetries). Higher-resolution imaging such as VLBI would be necessary to spatially disentangle the two components and constrain their morphologies independently.

\item \textbf{Uncertainty in source size estimates:} Physical parameter estimates depend critically on the angular size of the source. While we provide an indirect estimate, direct VLBI measurement of the source size would provide more accurate magnetic field and age results. The extended lobe is a particular case: it is unresolved, and imaging places no useful bound on it, because the lobe is faint at the frequencies where our angular resolution is highest and the beam is of order kiloparsecs at the frequencies where the lobe dominates. We therefore bound its size by expansion physics rather than by imaging (Section~\ref{sec:blobe}). Because $B_{\rm eq} \propto l^{-6/7}$, the resulting field is nonetheless constrained to within a factor of a few; a direct size measurement, from sensitive sub-arcsecond imaging at frequencies below ${\sim}2$\,GHz where the lobe contributes ${\gtrsim}50$ per cent of the flux, would tighten it considerably.

\item \textbf{Ambiguity in the absorption mechanism:} Our data cannot uniquely identify the physical origin of the low-frequency cutoff of the extended lobe. As discussed in Section~\ref{sec:jet_models}, both free-free absorption by circumnuclear ionised gas and an intrinsic low-energy cutoff in the relativistic electron population remain viable; we favour FFA, since the gas density and scale it requires match the warm absorber independently detected in X-rays \citep{Longinotti2009, Coffey2014}, but the present frequency coverage does not exclude the intrinsic-cutoff alternative, and the exponential cutoff we adopt is a phenomenological description rather than a specific mechanism. This ambiguity does not overturn our main conclusions --- the source is a compact, self-absorbed synchrotron source in the GPS/HFP regime regardless of the detailed origin of the lobe cutoff --- but it does affect some inferred parameters, such as the implied electron density or minimum Lorentz factor. Polarisation measurements and deeper sub-70\,MHz observations could help further constrain the absorption mechanism.

\end{itemize}

Despite these limitations, our broadband radio analysis has robustly identified a GPS/HFP-class spectral turnover in H\,0557--385 that is physically inconsistent with a canonical radio-quiet AGN, motivating further high-resolution and time-domain observations to determine the evolutionary fate of this intermediate-class source.

\section{Conclusions}
\label{sec:conclusions}

We have presented a broadband radio spectral analysis of H\,0557--385 ($z = 0.034$), a Seyfert\,1 galaxy conventionally classified as radio-quiet, combining new ATCA observations with archival MWA, ASKAP, and ALMA data to construct a broadband radio spectrum spanning 0.076--258.8\,GHz. Our principal findings and their implications are as follows.

\begin{enumerate}

\item \textbf{H\,0557--385 is not a canonical radio-quiet AGN.} The broadband radio spectrum exhibits a pronounced gigahertz-peaked turnover at ${\sim}9.4$\,GHz, requiring a two-component stratified synchrotron model: a compact self-absorbed component superposed on an extended power-law lobe. The radio-quiet classification of this source rests on a radio-to-optical flux-ratio metric that is insensitive to the physical state of the compact radio-emitting plasma. Evaluated on radio-physical grounds, H\,0557--385 is more consistent with a young or recently re-triggered radio galaxy than with a canonical radio-quiet Seyfert.

\item \textbf{The derived physical parameters place H\,0557--385 in the GPS/HFP regime.} The inferred source size (${\sim}15$\,pc), equipartition magnetic field (2.8\,mG), synchrotron spectral age ($\lesssim1.8\times10^3$\,yr), and jet power (${\sim}5\times10^{41}$--$5\times10^{44}$\,erg\,s$^{-1}$) are all characteristic of compact steep-spectrum (CSS) and gigahertz-peaked spectrum (GPS) young radio galaxies. The main distinguishing feature from classical GPS/HFP sources is the lower jet power, consistent with a low-luminosity young radio galaxy hosted by an S0a Seyfert rather than a massive elliptical.

\item \textbf{H\,0557--385 is a candidate missing link between radio-quiet Seyferts and classical young radio galaxies.} It occupies the interface between two traditionally distinct AGN populations: optically it resembles a Seyfert (accretion disc properties, host galaxy morphology, total luminosity), while radio-physically it has the spectral morphology and synchrotron properties of a young radio galaxy. This makes it a strong candidate for an intermediate object that bridges the evolutionary gap between the two classes.

\item \textbf{Two physically distinct interpretations remain viable.} Either (a) H\,0557--385 is a genuine young radio galaxy in the HFP/GPS phase that may evolve into a larger radio source if jet activity is sustained; or (b) it is a Seyfert AGN undergoing a brief episode of compact radio activity that will revert to a radio-quiet state once the current jet phase ends. Discriminating between these scenarios requires VLBI morphology, multi-epoch spectral monitoring, and polarisation measurements.

\item \textbf{Broadband surveys may uncover a missing population.} Sources of this kind, compact and self-absorbed with spectral peaks well above the standard classifying frequencies, are readily missed by single-frequency surveys. Broadband spectral coverage with facilities such as MeerKAT, ASKAP, the ngVLA, and eventually the SKA will be essential to establish how common they are, and H\,0557--385 motivates targeted searches for GHz-peaked spectra within optically classified radio-quiet samples.

\end{enumerate}

Looking forward, the following observations are needed to resolve the nature of H\,0557--385 and quantify the broader population of similar intermediate objects:
\begin{itemize}
\item \textbf{VLBI imaging} to directly resolve sub-parsec jet morphologies, measure (or constrain) proper motions, and confirm collimated plasma outflow as the origin of the compact emission.
\item \textbf{Multi-epoch broadband monitoring} (1--100\,GHz) to track the evolution of the spectral turnover frequency, which is predicted to migrate to lower frequencies as the source expands if it is in the genuine GPS/HFP phase.
\item \textbf{Linear polarisation and Faraday rotation} studies to probe magnetic field topology and, primarily, to test for beamed (blazar-like) contamination, since genuine young sources are expected to be weakly polarised; cleanly distinguishing the MAD versus CAD launching modes additionally requires sub-parsec (milliarcsecond-scale) resolution attainable only with high-frequency or space VLBI.
\item \textbf{Systematic broadband surveys} of optically radio-quiet AGN with ngVLA, MeerKAT, SKA, and ALMA to identify further missing-link objects and quantify the incidence of young-jet phases in Seyfert galaxies.
\item \textbf{GRMHD simulations} of low-power jet launching in disc-hosted, low-luminosity AGN to explore the conditions under which compact jets are produced, confined, and eventually quenched.
\end{itemize}

H\,0557--385 exemplifies a class of AGN that challenges our ability to classify radio activity from optical data alone. It is neither a typical radio-quiet Seyfert nor a classical radio-loud young radio galaxy, but rather an intermediate object that lies at the boundary between the two populations. Recognising and characterising this class will be essential for understanding the conditions that ignite and sustain radio AGN activity, for correctly estimating jet feedback across the AGN population, and for tracing the full evolutionary sequence from compact, newly ignited radio sources to large-scale radio galaxies.

\section*{Data Availability}
All raw data used in this study are available from public MWA (GLEAM--X; \citealt{HurleyWalker2022, Ross2024}), ASKAP, ATCA, and ALMA observatory archives.

\section*{Acknowledgements}
We thank the referee for their valuable comments. We thank Jamie Stevens and staff from the Australia Telescope National Facility for scheduling the ATCA radio observations. The ATCA is part of the Australia Telescope National Facility, which is funded by the Australian Government for operation as a National Facility managed by CSIRO. We acknowledge the Gomeroi people as the traditional owners of the ATCA Observatory site. This paper makes use of the following ALMA data: ADS/JAO.ALMA\#2016.1.01385.S. ALMA is a partnership of ESO (representing its member states), NSF (USA), and NINS (Japan), together with NRC (Canada), MOST and ASIAA (Taiwan), and KASI (Republic of Korea), in cooperation with the Republic of Chile. The Joint ALMA Observatory is operated by ESO, AUI/NRAO, and NAOJ. This scientific work uses data obtained from Inyarrimanha Ilgari Bundara, the CSIRO Murchison Radio-astronomy Observatory. We acknowledge the Wajarri Yamaji People as the Traditional Owners and Native Title Holders of the observatory site. Support for the operation of the MWA is provided by the Australian Government (NCRIS), under a contract to Curtin University administered by Astronomy Australia Limited. This research used the ALICE High Performance Computing facility at the University of Leicester. The authors acknowledge the use of Claude (Anthropic) in editing some sections of the manuscript for clarity and consistency. JC, RLCS acknowledge support from the Leverhulme Trust grant RPG-2023-240. FP acknowledges financial support from the Bando Ricerca Fondamentale INAF and ``Programma di Ricerca Fondamentale INAF 2023 and 2024''.

\bibliographystyle{mnras}
\bibliography{Manuscript_Final_Version}

\appendix

\section{Spectral Model Parameters and Posterior Distributions}

\begin{figure*}
    \includegraphics[width=2.2\columnwidth]{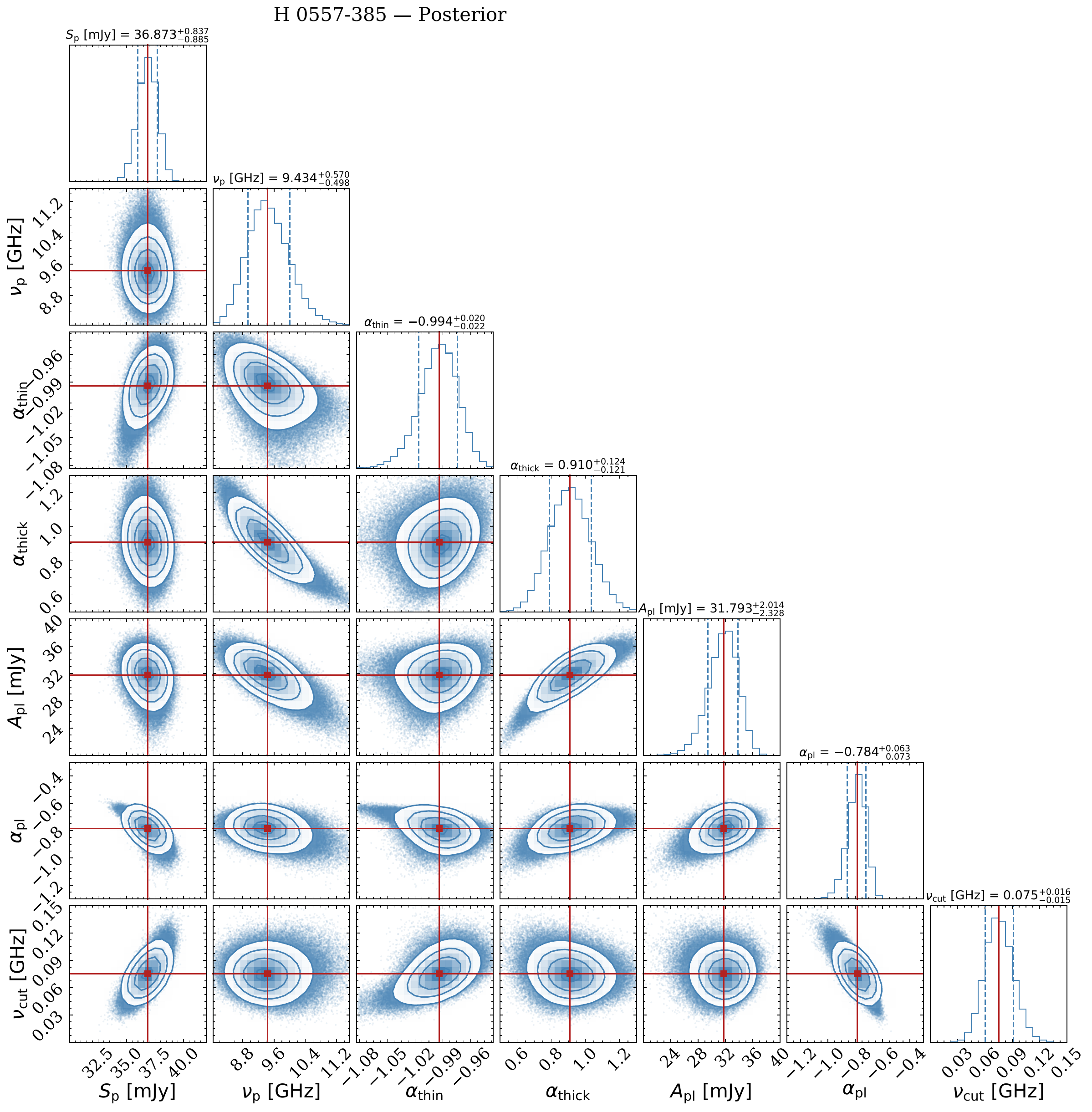}
    \caption{Corner plot illustrating the posterior probability distributions for the stratified jet model spectral parameters. Diagonal panels show marginal distributions; off-diagonal panels show joint distributions between parameter pairs. Dashed vertical lines indicate the 16th, 50th (median), and 84th percentiles. All parameters appear well constrained with tight contours and no significant degeneracies, supporting the robustness of the model.}
    \label{fig:figA1}
\end{figure*}

\begin{figure}
    \includegraphics[width=\columnwidth]{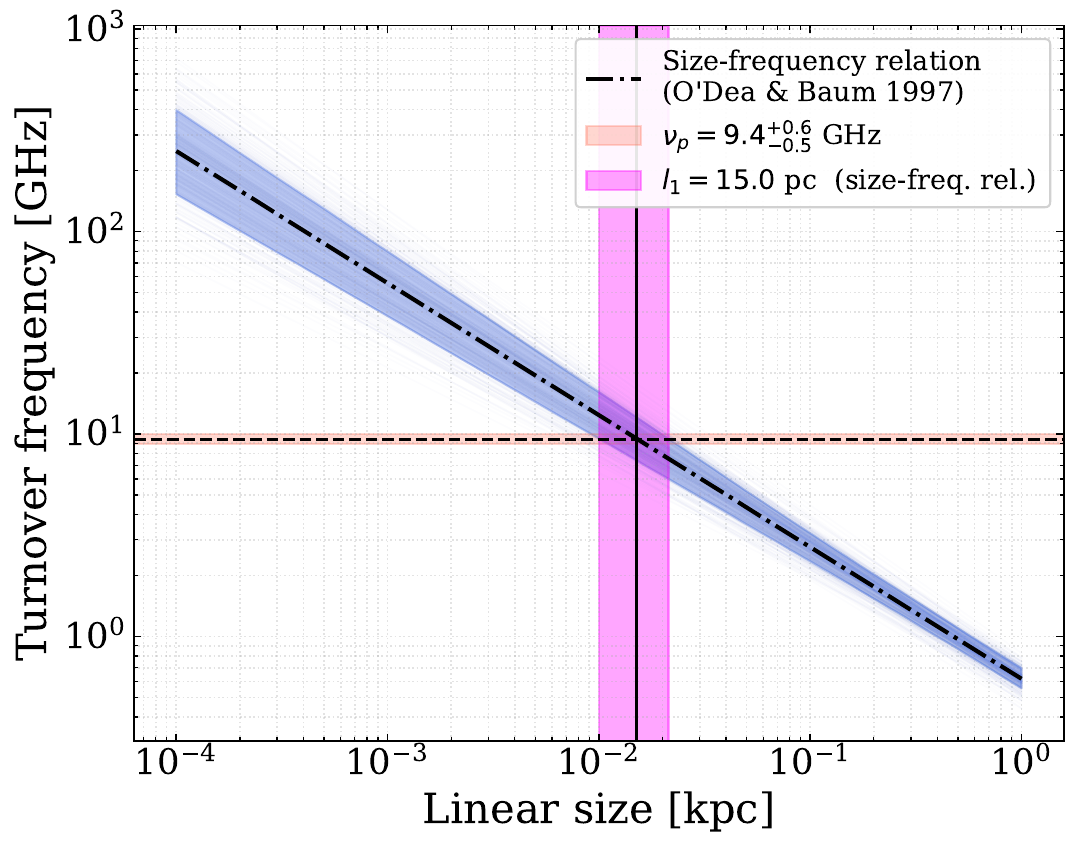}
    \caption{Empirical relation between the linear source size and the turnover frequency for synchrotron self-absorbed sources (Equation~\ref{turnover_size}), derived from GPS, CSS, and HFP sources \citep{Fanti1990, ODea1997, ODea1998, Keim2019}. The dot-dashed black line is the best-fit scaling relation; the blue shaded region shows its uncertainty. The horizontal dashed line marks the observed turnover frequency of H\,0557--385; the vertical solid line indicates the corresponding linear size. Shaded regions show the 16th--84th percentile ranges.}
    \label{fig:figA2}
\end{figure}

\begin{figure}
    \includegraphics[width=\columnwidth]{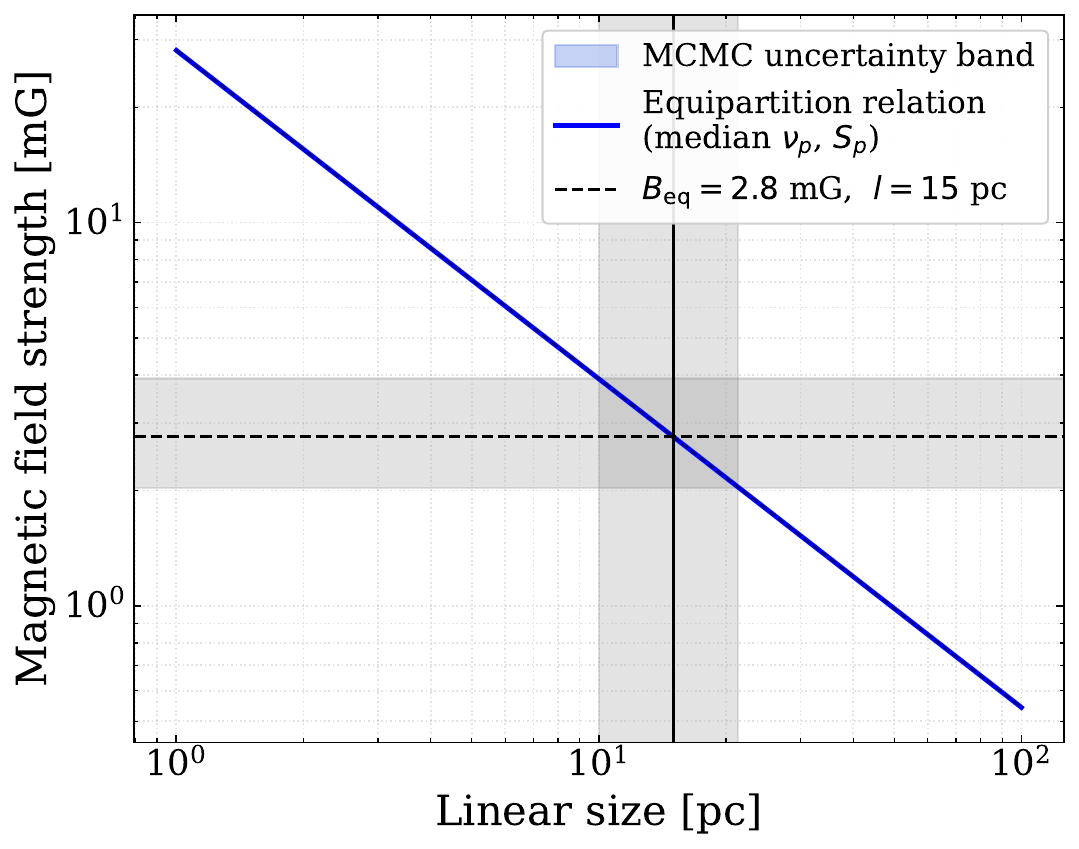}
    \caption{Empirical relation between linear source size and equipartition magnetic field strength following \citet{Miley1980} (Equation~\ref{eq:mag_field}). The solid blue line shows Equation~\ref{eq:mag_field} evaluated at the posterior medians of $\nu_{\rm p}$ and $S_{\rm p}$, and the blue band its 16th--84th percentile range over the MCMC draws. The vertical line marks the linear size of H\,0557--385 from the turnover--size relation and the horizontal dashed line the corresponding equipartition magnetic field strength; the shaded bands show their 16th--84th percentile ranges.}
    \label{fig:figA3}
\end{figure}

\begin{figure}
    \includegraphics[width=\columnwidth]{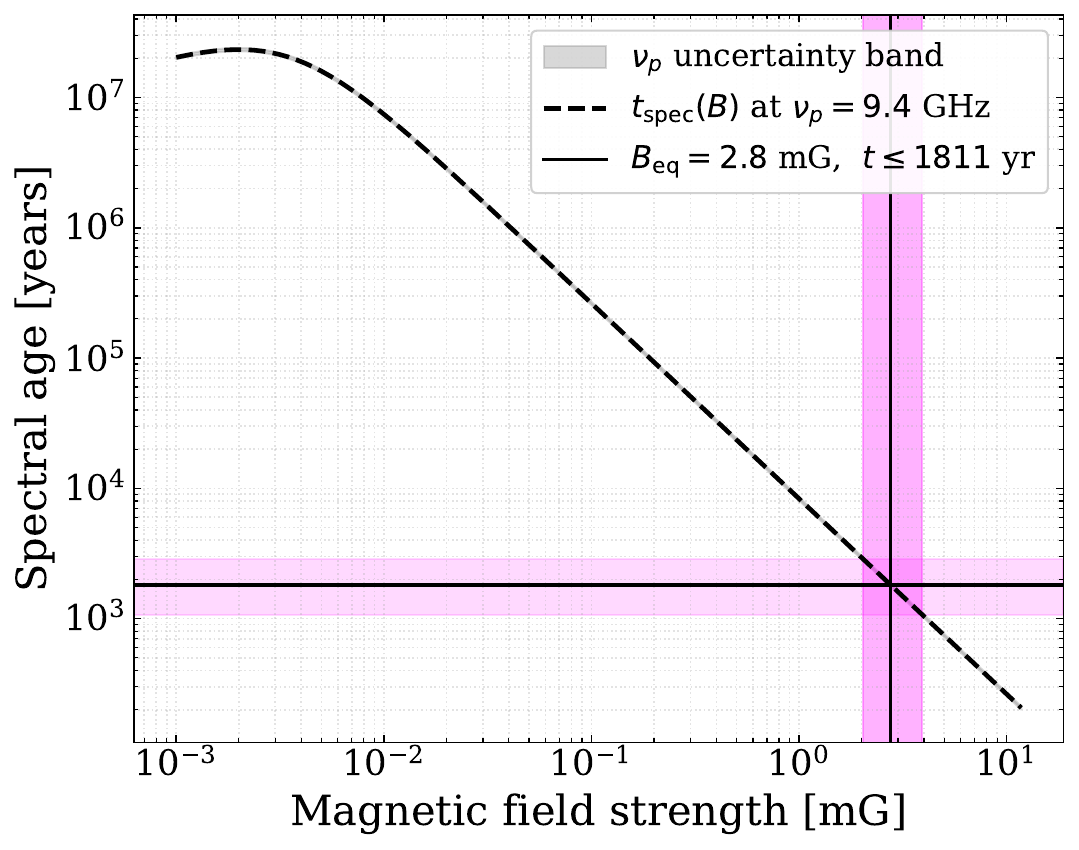}
    \caption{Synchrotron spectral age and magnetic field strength for compact, self-absorbed radio sources following \citet{Miley1980} (Equation~\ref{eq:spec_age}). The dashed curve shows Equation~\ref{eq:spec_age} at the median $\nu_{\rm p}$, and the grey band its 16th--84th percentile range over the MCMC draws of $\nu_{\rm p}$. The vertical line marks the equipartition magnetic field strength of H\,0557--385 and the horizontal line the corresponding spectral age; the shaded bands show their 16th--84th percentile ranges. The resulting upper limit, $\lesssim(1.8^{+1.1}_{-0.7})\times10^3$\,yr, is comparable to the ages of young GPS galaxies.}
    \label{fig:figA4}
\end{figure}

This appendix provides supporting material for the two-component spectral fit presented in Section~\ref{sec:jet_models}. Table~\ref{tab:fit_params} lists the prior range or distribution adopted for each of the seven free parameters together with its best-fit (posterior median) value and $1\sigma$ credible interval. Figure~\ref{fig:figA1} shows the full set of one- and two-dimensional marginal posterior distributions (the ``corner'' plot) for these parameters, from which the quoted uncertainties and any parameter degeneracies can be assessed. The remaining figures in this appendix (Figs~\ref{fig:figA2}--\ref{fig:figA4}) show the empirical scaling relations used in Section~\ref{sec:physical_params} to derive the source's physical properties.

\begin{table}
\centering
\caption{Priors and best-fit parameters for the two-component spectral model (Section~\ref{sec:jet_models}). Best-fit values are posterior medians with $1\sigma$ credible intervals from the MCMC analysis. Gaussian priors were centred on the results of an initial simultaneous fit to the same data; the remaining parameters used uniform priors over the stated range.}
\label{tab:fit_params}
\scalebox{1.0}{%
\begin{tabular}{lccc}
\hline
\textbf{Parameter} & \textbf{Prior type} & \textbf{Prior range } & \textbf{Best-fit value} \\
 &  & \textbf{/ distribution} & \\
\hline
$S_{\rm p}$ (mJy)          & Gaussian & $\mathcal{N}(40.1,\,5.9)$      & $36.9^{+0.8}_{-0.9}$ \\
$\nu_{\rm p}$ (GHz)        & Gaussian & $\mathcal{N}(10.7,\,4.0)$     & $9.4^{+0.6}_{-0.5}$ \\
$\alpha_{\rm thin}$        & Gaussian & $\mathcal{N}(-1.01,\,0.08)$   & $-0.99\pm0.02$ \\
$\alpha_{\rm thick}$       & Uniform  & $[0,\,2.5]$                   & $+0.91\pm0.12$ \\
$A_{\rm pl}$ (mJy)         & Uniform  & $[1,\,60]$                    & $31.8^{+1.9}_{-2.3}$ \\
$\alpha_{\rm pl}$          & Uniform  & $[-2.2,\,-0.3]$               & $-0.78^{+0.06}_{-0.07}$ \\
$\nu_{\rm cut}$ (MHz)      & Gaussian & $\mathcal{N}(43.5,\,45.4)$    & $75^{+16}_{-15}$ \\
\hline
\end{tabular}}
\end{table}

\bsp	
\label{lastpage}
\end{document}